\documentclass[11pt]{article}
 \usepackage{titlesec}
\usepackage{hyperref}

\titleclass{\subsubsubsection}{straight}[\subsection]

\usepackage[utf8]{inputenc}
\usepackage{amsmath,amsfonts,amssymb,stackengine,graphicx}

\usepackage[utf8]{inputenc}
\newif\iffigs\figstrue

\usepackage[title]{appendix}
\usepackage{epsfig,latexsym}
\usepackage{hyperref}
\usepackage{amsmath}
\usepackage{verbatim}
\usepackage{color}
\usepackage{mathrsfs}
\usepackage{slashed}
\usepackage{amssymb}

\DeclareMathAlphabet{\mathpzc}{OT1}{pzc}{m}{it}

 \csname
@addtoreset\endcsname{equation}{section}

\def\gz0{\gamma^{0}}

\def\m{\mu}

\def\beq{\begin{equation}}
\def\eeq{\end{equation}}
\def\bea{\begin{eqnarray}}
\def\eea{\end{eqnarray}}
\def\ba{\begin{array}}
\def\ea{\end{array}}
\def\bec{\begin{center}}
\def\ec{\end{center}}
\def\ba{\begin{align}}
\def\ena{\end{align}}

\def\12{\frac{1}{2}}

\newcounter{subsubsubsection}[subsubsection]
\renewcommand\thesubsubsubsection{\thesubsubsection.\arabic{subsubsubsection}}

\titleformat{\subsubsubsection}
  {\normalfont\normalsize\bfseries}{\thesubsubsubsection}{1em}{}
\titlespacing*{\subsubsubsection}
{0pt}{3.25ex plus 1ex minus .2ex}{1.5ex plus .2ex}

\makeatletter
\renewcommand\paragraph{\@startsection{paragraph}{5}{\z@}%
  {3.25ex \@plus1ex \@minus.2ex}%
  {-1em}%
  {\normalfont\normalsize\bfseries}}
\renewcommand\subparagraph{\@startsection{subparagraph}{6}{\parindent}%
  {3.25ex \@plus1ex \@minus .2ex}%
  {-1em}%
  {\normalfont\normalsize\bfseries}}
\def\toclevel@subsubsubsection{4}
\def\toclevel@paragraph{5}
\def\toclevel@paragraph{6}
\def\l@subsubsubsection{\@dottedtocline{4}{7em}{4em}}
\def\l@paragraph{\@dottedtocline{5}{10em}{5em}}
\def\l@subparagraph{\@dottedtocline{6}{14em}{6em}}
\makeatother

\begin{document}

\begin{flushright}
{\today}
\end{flushright}

\vspace{10pt}

\begin{center}


{\Large\sc Non--Supersymmetric Vacua and Self--Adjoint Extensions}


\vspace{25pt}
{\sc J.~Mourad${}^{\; a}$  \ and \ A.~Sagnotti${}^{\; b}$\\[15pt]

${}^a$\sl\small APC, UMR 7164-CNRS, Universit\'e   Paris Cit\'e  \\
10 rue Alice Domon et L\'eonie Duquet \\75205 Paris Cedex 13 \ FRANCE
\\ e-mail: {\small \it
mourad@apc.univ-paris7.fr}\vspace{10
pt}

{${}^b$\sl\small
Scuola Normale Superiore and INFN\\
Piazza dei Cavalieri, 7\\ 56126 Pisa \ ITALY \\
e-mail: {\small \it sagnotti@sns.it}}\vspace{10pt}
}

\vspace{40pt} {\sc\large Abstract}\end{center}
\noindent
Internal intervals spanned by finite ranges of a conformal coordinate $z$ and terminating at a pair of singularities are a common feature of many string compactifications with broken supersymmetry. The squared masses emerging in lower--dimensional Minkowski spaces are then eigenvalues of Schr\"odinger--like operators, whose potentials have double poles at the ends of the intervals. For one--component systems, the possible self--adjoint extensions of Schr\"odinger operators are described by points in $AdS_3 \times S^1$, and those corresponding to independent boundary conditions at the ends of the intervals by points on the boundary of $AdS_3$. The perturbative stability of compactifications to Minkowski space time depends, in general, on these choices of self--adjoint extensions. We apply this setup to the orientifold vacua driven by the ``tadpole potential'' $V=T \ e^{\,\frac{3}{2}\,\phi}$ and find, in nine dimensions, a massive scalar spectrum, a unique choice of boundary conditions with stable tensor modes and a massless graviton, and a wide range of choices leading to massless and/or massive vector modes.
\vskip 12pt

\setcounter{page}{1}

\pagebreak

\newpage
\tableofcontents
\newpage
\baselineskip=20pt
\section{\sc  Introduction and Summary}\label{sec:intro}

This work is devoted to eliciting the possible self--adjoint boundary conditions for one--dimensional Schr\"odinger systems defined in finite intervals and  their implications for string vacua with broken supersymmetry. We were repeatedly confronted with aspects of this mathematical problem, which has a long history (see, for example,~\cite{math_literature_1,math_literature_2}) when studying the modes present in the warped compactifications found in~\cite{ms21_1,ms21_2}. These backgrounds are string--inspired solutions of the supergravity~\cite{supergravity} effective action, which can include the ``tadpole potentials''
\beq
V \ = \ T \ e^{\gamma\,\phi} \label{tadpole_intro}
\eeq
that accompany the breaking of supersymmetry in String Theory~\cite{strings}. They are generalizations of the Dudas--Mourad vacuum of~\cite{dm_vacuum}, and share with it the emergence of a finite internal interval. An internal interval had previously made its entry, in String Theory, in the Horava--Witten recovery~\cite{hw} of the $E_8\times E_8$ heterotic string~\cite{heterotic} from the eleven--dimensional supergravity of Cremmer, Julia and Scherk~\cite{cjs}. However, the examples that motivated the present analysis contain an additional ingredient, supersymmetry breaking, which can substantially affect vacuum stability. Consequently, one cannot forego a careful scrutiny of the resulting modes, which depend to a crucial extent, as we shall see, on the possible choices of self--adjoint extensions.

The tadpole potential is the leading back-reaction to the breaking of supersymmetry in the three non--tachyonic string models of~\cite{so1616,as95,sugimoto}. The first is a heterotic model, where the contribution in eq.~\eqref{tadpole_intro} emerges from the torus amplitude and $\gamma=\frac{5}{2}$, while the others are orientifolds~\cite{orientifolds}, where the potential emerges at the (projective) disk level and the parameter $\gamma$ has the ``critical'' value $\gamma_c=\frac{3}{2}$. The last model embodies the simplest realization of ``brane supersymmetry breaking''~\cite{bsb}, a peculiar mechanism induced by the simultaneous presence, in the vacuum, of extended BPS objects, anti-branes and orientifolds, preserving incompatible portions of the supersymmetries originally present in ten--dimensional Minkowski space. These extended objects result in the complete breaking of supersymmetry at the string scale, which appears non--linearly realized in the low--energy supergravity~\cite{dm_gravitino}. In ten dimensions there are no tachyonic modes, but there is also no order parameter to recover an unbroken phase.

The non--trivial profiles considered in~\cite{ms21_1,ms21_2} for the metric, the dilaton and a $p+1$-form potential describe compactifications to lower--dimensional Minkowski spaces, with intervals of finite length that are covered by finite spans $0 \leq z \leq z_m$ of a conformal coordinate, and the resulting modes can be associated to Schr\"odinger--like systems after suitable field redefinitions. The possible choices of boundary conditions at the ends of the interval are thus essential to characterize the resulting spectra.
In~\cite{bms,ms22_1} we discussed Hermitian Schr\"odinger systems for Bose fields of the type
 \beq
 {H}\, \psi \ = \ m^2\, \psi \,  \label{schr_intro}
 \eeq
and their counterparts for Fermi systems, but we did not address in detail the possible choices of self--adjoint extensions. Self--adjoint boundary conditions grant the reality of the eigenvalues and the completeness of the resulting modes, which is instrumental to address all possible origins of instability, but they do not exclude, in general, the presence of negative eigenvalues. In reductions to lower--dimensional Minkowski spaces instabilities manifest themselves precisely, at the perturbative level, as \emph{negative} $m^2$ eigenvalues of the Schr\"odinger systems~\eqref{schr_intro}, and the relevance of the sign is precisely where our problem departs from the conventional setup. As we shall see, negative eigenvalues can emerge even in cases where, at first sight, one would be inclined to exclude them. As a result, vacuum stability can only hold, in general, with special self--adjoint boundary conditions granting positive spectra for the Schr\"odinger operators~\eqref{schr_intro}, among those compatible with the symmetries of the backgrounds.

The plan of this paper is as follows. In Section~\ref{sec:sing_pot} we motivate the appearance of Schr\"odinger problems with double--pole singularities at the ends of the intervals of interest for the types of warped backgrounds encountered originally in~\cite{dm_vacuum}, and more recently in~\cite{ms21_1,ms21_2}. In Section~\ref{sec:selfadjoint} we discuss in detail how to characterize self--adjoint extensions for one--dimensional Schr\"odinger systems in a finite interval. We begin from the free theory, thus impinging on old results that have been widely considered in Mathematics~\cite{math_literature_1,math_literature_2}, but from a perspective that seems to us potentially useful. We then turn to the types of Hamiltonians of direct interest to us, with double poles at the ends of an interval, and we discuss in detail a class of exactly solvable cases of this type. As we shall see, while the modes resulting from singular potentials typically develop algebraic singularities at the ends of the interval, one can still characterize the allowed self--adjoint boundary conditions along the lines of non--singular problems. In Section~\ref{sec:applications} we elaborate on some lessons of the analysis for the actual vacua of~\cite{dm_vacuum,ms21_1,ms21_2}. We collect our conclusions in Section~\ref{sec:conclusions}.

\section{\sc  Singular Potentials from Warped String Vacua}\label{sec:sing_pot}

The solutions discussed in~\cite{ms21_1,ms21_2} can clearly motivate the analysis presented in this paper, and conversely the present analysis can complement the previous work on their modes in~\cite{bms}. To this end, we can actually confine our attention to the isotropic nine--dimensional solutions that can be deduced starting from the Einstein--frame effective action
\beq
{\cal S} \ = \ \frac{1}{2\,k_{10}^2} \ \int \ d^{10}x \ \sqrt{-g} \left( R \ - \ \frac{1}{2}\,\partial^M\,\phi\,\partial_M \,\phi  \ - \ T \, e^{\gamma\,\phi} \right) \ . \label{Taction}
\eeq
These generalize the results in~\cite{dm_vacuum} and share with them some features that are central to our analysis, but nonetheless are relatively simple to characterize.
Abiding to common practice, we have already called the last term in eq.~\eqref{Taction} ``tadpole potential'', while paying attention the ``critical'' value
\beq
\gamma_c \ = \ \frac{3}{2} \ .
\eeq
This characterizes the non--tachyonic ten--dimensional orientifold models with broken supersymmetry and will play an important role in the following.

In~\cite{ms21_1} we found two isotropic nine--dimensional solutions that follow from the action~\eqref{Taction} for $T=0$, for which
\beq
ds^2 \ = \  \left(\mu_0\,\xi\right)^\frac{2}{9} \, dx^2 \ + \ d \xi^2  \ , \qquad
e^\phi \ = \  \left(\mu_0\,\xi\right)^{\,\pm\,\frac{4}{3}} \, e^{\phi_0} \ , \label{spontaneous_r_isotropic}
\eeq
where $\xi>0$ and $\mu_0$ and $\phi_0$ are arbitrary constants. These two solutions are related to one another by the redefinition $\phi \to - \phi$, which is a manifest symmetry of the action for $T=0$. In both cases, the interval has an infinite length and the string coupling diverges, as $\xi \to \infty$ or as $\xi \to 0$.

When $T \neq 0$ the solutions depend on $\gamma$, which we can assume to be larger than zero, up to a redefinition of $\phi$, but one must distinguish different ranges for it~\cite{ms21_2}. Still, the $T=0$ solutions~\eqref{spontaneous_r_isotropic} continue, surprisingly, to play a role in asymptotic regions, as we can now recall.

\begin{itemize}
\item For $\gamma=\gamma_c$ the solution is the one originally found in~\cite{dm_vacuum}, and reads
\beq
ds^2 \ = \  e^{\,-\,\frac{u}{6}} \, {u}^\frac{1}{18}\, dx^2 \,+\, \frac{2}{3\,T\,u^\frac{3}{2}}\ e^{\,-\,\frac{3}{2}\left(u+\phi_0\right)}\, {du^2} \ , \qquad
e^\phi \ = \  e^{\,u\,+\,{\phi}_0}\, {u}^\frac{1}{3} \ . \label{dmsol}
\eeq
Here $u \geq 0$, and now the internal interval has a finite length. However, the additional substitution
\beq
\mu_0\,\xi \ \sim \ u^\frac{1}{4} \ ,
\eeq
where $\mu_0$ depends on $T$ and $\phi_0$,
shows that, up to a rescaling of the spatial coordinates, the limiting behavior in the vicinity of the boundary at $u=0$ is dominated by
\beq
ds^2 \ \sim \  \left(\mu_0\,\xi\right)^\frac{2}{9} \, dx^2 \,+\, {d\xi^2} \ , \qquad
e^\phi \ \sim \ \left(\mu_0\,\xi\right)^{\frac{4}{3}} \ . \label{limit_l}
\eeq

This is a first instance of the role played by the solutions~\eqref{spontaneous_r_isotropic} even for $T \neq 0$, as we had anticipated. In this first case the correspondence is perhaps not too surprising, since the string coupling tends to zero as $\xi \to 0$, but something similar also occurs, strikingly, for large values of $u$. In this limit the powers have negligible effects compared to the exponential terms, and the background in eqs.~\eqref{dmsol} approaches
\beq
ds^2 \ = \  e^{\,-\,\frac{u}{6}} \, dx^2 \,+\, \frac{1}{T}\ e^{\,-\,\frac{3}{2}\, u} {du^2} \ , \qquad
e^\phi \ = \  e^{\,u} \ . \label{dmsol2}
\eeq
One can now let
\beq
\mu_0\left(\xi_m \ - \ \xi\right) \ \sim \ e^{\,-\,\frac{3}{4}\, u} \ ,
\eeq
where $\mu_0$ is determined by $T$, so that $\xi_m$ measures the total proper length of the interval, with
\beq
\mu_0\,\xi_m \ = \ \frac{2}{\sqrt{T}}\ \left(3\,e^{\,\phi_0}\right)^{\,-\,\frac{3}{4}}\, \Gamma\left(\frac{1}{4}\right) \ .
\eeq
In the neighborhood of the second boundary at $\xi_m$, up to a rescaling of the spatial coordinates, the background thus approaches
\beq
ds^2 \ \sim \  \left[\mu_0\left(\xi_m\ - \ \xi\right)\right]^\frac{2}{9} \, dx^2 \,+\, {d\xi^2} \ , \qquad
e^\phi \ \sim \ \left[\mu_0\left(\xi_m\ - \ \xi\right)\right]^{-\frac{4}{3}} \ . \label{limit_r}
\eeq
This limiting behavior is also captured by one of the isotropic \emph{tensionless} solutions in eq.~\eqref{spontaneous_r_isotropic} that is now approached, surprisingly, within a region of strong coupling. The tadpole ought to dominate at this end of the interval. Yet, while it is instrumental to grant the internal interval a finite length and determines the scale $\mu_0$, it has somehow negligible effects on the asymptotic structure of the background.

\item For $\gamma<\gamma_c$ the solutions read
\bea
ds^2 &=& e^{\,-\, \frac{2\,\gamma\,\lambda\,r}{\gamma_c}}\ \frac{dx^2}{\left[\Delta\,\cosh\left({r}\right)\right]^{2\,\lambda}}  \ +\ \frac{e^{\,-\,\gamma\left(\frac{18 \,\lambda}{\gamma_c}\,r\,+\,\phi_0\right)}\ dr^2}{\left[\Delta\,\cosh\left({r}\right)\right]^{18\,\lambda}} \ , \nonumber \\
e^{\phi} &=&  \left[\Delta\,\cosh\left({r}\right)\right]^{\frac{18\,\gamma}{\gamma_c^2}\,\lambda} \ e^{\frac{18 \,\lambda}{\gamma_c}\,r\,+\,\phi_0} \ ,
\eea
where now $- \infty < r < + \infty$,
\beq
\Delta^2 \ = \  \frac{T}{2}\ \left|\gamma^2\,-\,\gamma_c^2\right| \ , \qquad \lambda \ = \ \frac{1}{9\left(1 \ -\ \frac{\gamma^2}{\gamma_c^2} \right)} \label{lambda_sub}
\eeq
and $\phi_0$ is a constant. This class of solutions describes again compactifications on intervals of finite length, where the string coupling vanishes at one end and diverges at the other. Still, the reader can verify that, as for $\gamma=\gamma_c$, the limiting behavior at both ends is captured by the isotropic tensionless solutions in eqs.~\eqref{limit_l} and \eqref{limit_r}, albeit with different finite values of $\xi_m$.

\item {For $\mathbf{\gamma} \,\mathbf{>}\, \mathbf{\gamma}_\mathbf{c}$} there are altogether three classes of solutions. The solutions in the first class read
\beq
ds^2 \ = \  r^{2|\lambda|}\ dx^2  \,+\, \frac{r^{18\,|\lambda|}}{\Delta^2}\, e^{\,-\,\gamma\,\phi_0}\ dr^2 \ , \qquad
e^{\phi} \ = \   r^{\,-\,\frac{18\,\gamma}{\gamma_c^2}\,|\lambda|} \ e^{\phi_0} \ ,
\eeq
where $0 < r < \infty$.
They describe intervals of infinite length, and the string coupling diverges at the origin. Letting
\beq
\Delta\ \xi \ = \ e^{\,-\,\frac{\gamma}{2}\,\phi_0}\ \frac{\left( r\right)^{9\,|\lambda| +1}}{\left(9\,|\lambda| +1\right)} \ ,
\eeq
the preceding expressions become, for $0 < \xi < \infty$,
\beq
ds^2 \ = \ \left(\Delta\ \xi\right)^{\frac{2 \,\gamma_c^2}{9\,\gamma^2}}\ dx^2  \ +\  {d\xi^2} \ , \qquad
e^{\phi} \ = \   \left(\Delta \ \xi\right)^{\,-\,\frac{2}{\gamma}} \ e^{\phi_0} \ , \label{near0e0}
\eeq
after absorbing some multiplicative constants in rescalings of the $x$ coordinates and in redefinitions of $\phi_0$. Within this range matters are different and, in contrast with what happens for $\gamma \leq \gamma_c$, the limiting behavior at both ends is no longer captured by the tension--free solutions~\eqref{spontaneous_r_isotropic}.

In addition, there are the two branches of solutions
\bea
ds^2 &=& e^{\,\mp\, \frac{2\,\gamma\,|\lambda|\,r}{\gamma_c}}\ {\left[\Delta\,\sinh\left({r}\right)\right]^{2\,|\lambda|}} \  dx^2
  \ + \  \left[\Delta\,\sinh\left({r}\right)\right]^{18\,|\lambda|} \ e^{\,\mp\, \frac{18\,\gamma\,|\lambda|\,r}{\gamma_c}}\ e^{\,-\,\gamma\,\phi_0}\ dr^2 \ , \nonumber \\
e^{\phi} &=& \ \frac{e^{\,\phi_0}\ e^{\,\pm\, \frac{18\,|\lambda|\,r}{\gamma_c}}}{\left[\Delta\,\sinh\left({r}\right)\right]^{\frac{18\,\gamma}{\gamma_c^2}\,|\lambda|} } \ , \label{isotropicepos}
\eea
where $0 < r < \infty$. In both cases, the string coupling vanishes as $r \to \infty$ but diverges at $r=0$. Moreover, the length of the $r$-interval is finite in the first branch (upper sign), while it is infinite in the second (lower sign).
Close to $r=0$ the string coupling is unbounded and the limiting behavior is as in eqs.~\eqref{near0e0}. On the other hand, as $r\to \infty$ the string coupling vanishes and the limiting behavior is captured once more, for the two branches, by the isotropic tensionless solutions~\eqref{spontaneous_r_isotropic}.
\end{itemize}

A special presentation of the metrics will play a prominent role in the following sections, since it is closely related to Schr\"odinger systems. It rests on the introduction of conformal coordinates, which turn them into the form
\beq
ds^2 \ = \ e^{2\Omega(z)}\left(dx^2 \ + \ dz^2 \right) \ .
\eeq
For $\gamma=\gamma_c$ the conformal coordinate is defined as
\beq
z(r) \ = \ \int_0^r du \ \sqrt{\frac{2}{3\ T}}\ e^{\,-\,\frac{2}{3}\,u\,-\,\frac{3}{4}\,\phi_0} \ u^{\,-\,\frac{7}{9}}  \ ,
\eeq
while for $\gamma < \gamma_c$ it is defined as
\beq
z(r) \ = \ \int_{-\,\infty}^{r} \frac{du}{\left(\Delta\,\cosh u \right)^{8\lambda}} \ e^{\,-\,\gamma\left(\frac{8\lambda u}{\gamma_c}\ +\ \frac{\phi_0}{2}\right)} \ .
\eeq
In both cases $z$ has finite range $\left(0\leq z \leq z_m\right)$, with different values for $z_m=z(\infty)$. The limiting behavior close to the left end of the interval is now
\beq
ds^2 \ \sim \ \left(\frac{8 \mu_0 z}{9}\right)^\frac{1}{4}\left(dx^2 \ + \ dz^2 \right) \ , \qquad
e^\phi \ \sim \ \left(\frac{8 \mu_0 z}{9}\right)^\frac{3}{2} \ ,
\eeq
while close to the right end
\beq
ds^2 \ \sim \ \left[\frac{8 \mu_0 \left(z_m-z\right)}{9}\right]^\frac{1}{4}\left(dx^2 \ + \ dz^2 \right) \ , \qquad
e^\phi  \ \sim \  \left[\frac{8 \mu_0 \left(z_m-z\right)}{9}\right]^{-\,\frac{3}{2}} \ .
\eeq

For $\gamma > \gamma_c$, as we have seen, there are three classes of solutions. Only those belonging to the second class, which live in an interval of finite length, and for which
the conformal coordinate $z$ can be defined as
\beq
z(r) \ = \ \int_{0}^{r} \  du \ {\left(\Delta\,\sinh u \right)^{8\left|\lambda\right|}} \ e^{\,-\,\gamma\left(\frac{8\left|\lambda\right| u}{\gamma_c}\ +\ \frac{\phi_0}{2}\right)} \ ,
\eeq
lead to a finite value for $z_m = z(\infty)$.
Now close to $z=0$
\beq
ds^2 \ \sim \  z^\frac{2\,\gamma_c^2}{9\,{\gamma}^2 \ - \ {\gamma}_c^2} \left(dx^2 \ + \ dz^2 \right) \ , \qquad
e^\phi \ \sim \   z^{\,-\,\frac{18 \,\gamma}{9\,{\gamma}^2 \ - \ \gamma_c^2}} \ ,
\eeq
while close to other end of the interval, which lies at $z=z_m$,
\beq
ds^2 \ \sim \ \left(\frac{8 \mu_0 \left(z_m \ - \ z\right)}{9}\right)^\frac{1}{4}\left(dx^2 \ + \ dz^2 \right) \ , \qquad
e^\phi \ \sim \  \left(\frac{8 \mu_0 \left(z_m \ - \ z\right)}{9}\right)^\frac{3}{2} \ .
\eeq

Following~\cite{bms}, one can define tensor and scalar perturbations of these backgrounds according to
\beq
ds^2 \ = \  e^{2\,\Omega(z)} \Big[ \eta_{MN} \ + \ h_{MN}(x) f(z)\Big] dx^M \, dx^N \ , \qquad \phi \ = \ \phi(z) \ + \ \varphi(x)\,g(z) \ ,
\eeq
and there are also perturbations of the Yang--Mills fields and of the Ramond--Ramond two--form potential. The last two sets of modes can be conveniently described in a ``radial'' gauge, letting
\beq
A_z\ \ = \ 0 \ , \qquad B_{z M} \ = \ 0 \ .
\eeq

One is thus led to Schr\"odinger--like equations, with Hamiltonians
\beq
H \ = \ - \ \partial_z^{\,2} \ + \ V(z) \label{ham}
\eeq
that can be cast in the form
\beq
{H} \ = \ b \ + \ {\cal A}\,{\cal A}^\dagger \ ,
\eeq
where
\beq
{\cal A} \ = \ \partial_z \ +\ \frac{a}{2} \ , \qquad  {\cal A}^\dagger \ = \ - \ \partial_z \ +\ \frac{a}{2} \ ,
\eeq
so that the Schr\"odinger potential is
\beq
V \ = \ b \ + \ \frac{1}{2}\, a'\ + \ \frac{1}{4}\, a^2 \ , \label{V_schrod}
\eeq
where the ``prime'' indicates a derivative with respect to $z$.
In detail, for tensor perturbations
\beq
a \ = \ 8 \,\Omega' \ , \qquad b \ = \ 0 \ ,
\eeq
while for scalar perturbations~\footnote{There is a subtlety, explained in~\cite{bms}, when using the first type of expressions as $\gamma \to \gamma_c$.}
\bea
a &=& 24\,\Omega' \ - \ \frac{2 \ T\ \gamma}{\phi'} \ e^{2\Omega\,+\,\gamma\,\phi} \ = \  24 \, \Omega' \ - \ \frac{\gamma}{\phi'}\left[\left(\phi'\right)^2 \ - \ \left(12\,\Omega'\right)^2\right] \ , \nonumber \\
b &=& \frac{7}{4}\, T\, e^{2\,\Omega\,+\,\gamma\,\phi}\left(1\ + \ 8 \, \gamma \ \frac{\Omega'}{\phi'}\right) \ = \ \frac{7}{8\,\phi'}\left[\left(\phi'\right)^2 \ - \ \left(12\,\Omega'\right)^2\right]\left[ \phi'\ + \ 8\,\gamma\,\Omega'\right]\ . \label{ab}
\eea
Using the explicit forms of the background, one can show that $b>0$ for the three cases where $z_m$ is finite, while $b<0$ in the other two cases. To this end, it is convenient to use the first expression in the second of eqs.~\eqref{ab}.

For the orientifolds of~\cite{as95,sugimoto}, with $\gamma=\gamma_c$, there are also vector and two--form perturbations, whose Schr\"odinger potentials have again $b=0$ and
\bea
a_V &=& 8\,\Omega' \ + \ \frac{1}{2}\,\phi' \ , \nonumber \\
a_{RR} &=& 6\,\Omega' \ + \ \phi' \ .
\eea
On the other hand, for the $SO(16)\times SO(16)$ heterotic model, with $\gamma=\frac{5}{2}$, the vector and two--form perturbations have once more $b=0$, but
\bea
a_V &=& 8\,\Omega' \ - \ \frac{1}{2}\,\phi' \ , \nonumber \\
a_{NS} &=& 6\,\Omega' \ - \ \phi' \ .
\eea

The Schr\"odinger wavefunctions $\Psi$ for the two cases of tensor and scalar perturbations are proportional to the two functions $f(z)$ and $g(z)$, and the proportionality factor is
\beq
e^{\,-\,\frac{1}{2}\, \int dz \ a(z)} \ .
\eeq
Similar steps can be followed for the other types of bosonic perturbations.
\begin{figure}[ht]
\centering
\begin{tabular}{cc}
\includegraphics[width=55mm]{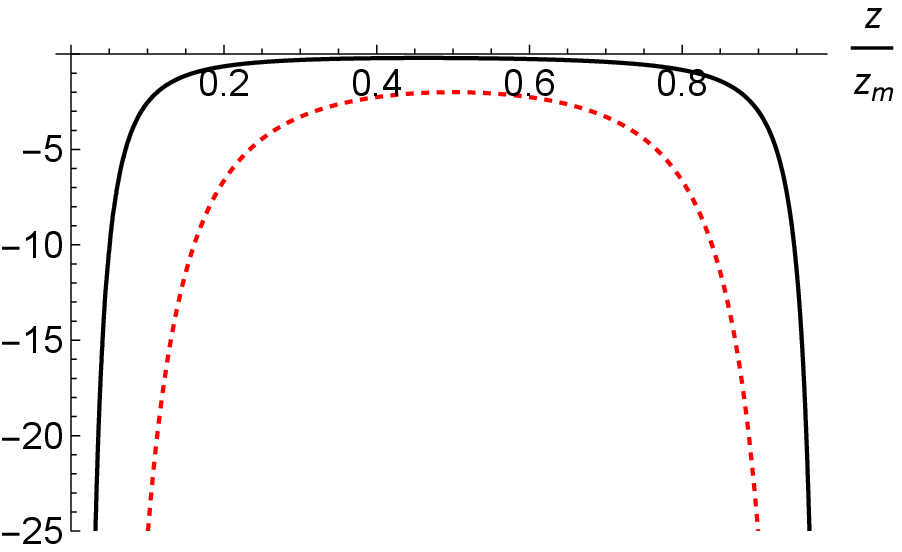} \qquad \qquad &
\includegraphics[width=55mm]{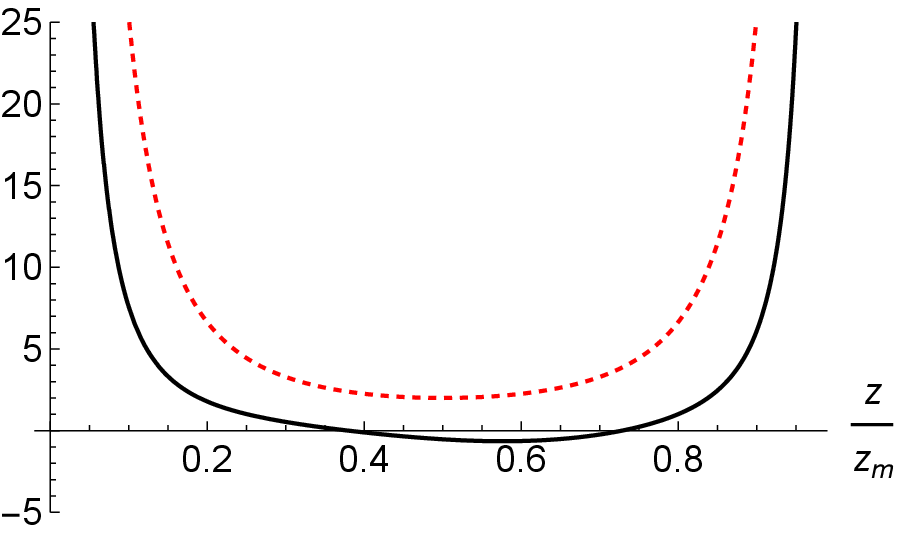} \\
\end{tabular}
\caption{\small The potentials for tensor perturbations (left panel, solid) and for scalar perturbations (right panel, solid), in units of $\frac{1}{z_m^2}$, for $\gamma=\gamma_c$, as functions of $\frac{z}{z_m}$, and their approximations obtained adding the two limiting behaviors in eq.~\eqref{limiting} (dashed).}
\label{fig:xi_quantized}
\end{figure}

The Schr\"odinger potentials~\eqref{V_schrod} for the different perturbations develop double--pole singularities at the ends of the interval, and in their neighborhood they behave as
\beq
V \ \sim \ \frac{\mu^2 \ - \ \frac{1}{4}}{z^2} \ , \qquad  V \ \sim \ \frac{\tilde{\mu}^2 \ - \ \frac{1}{4}}{\left(z_m \ - \ z\right)^2} \ . \label{limiting}
\eeq

\begin{itemize}
\item For $\gamma \leq \gamma_c$, the preceding expressions imply that for tensor perturbations
\beq
a \ = \ \frac{1}{z} \ , \qquad b \ = \ 0
\eeq
close to the origin, so that eq.~\eqref{V_schrod} implies that
\beq
\mu \ = \ 0 \ ,
\eeq
and
\beq
a \ = \ - \ \frac{1}{z_m \ - \ z} \ , \qquad b \ = \ 0
\eeq
close to $z_m$, so that eq.~\eqref{V_schrod} implies that
\beq
\tilde{\mu} \ = \ 0 \ .
\eeq
For scalar perturbations
\beq
a \ = \ \frac{3}{z} \ , \qquad b \ = \ 0
\eeq
close to the origin, so that eq.~\eqref{V_schrod} implies that
\beq
\mu \ = \ 1 \ ,
\eeq
and
\beq
a \ = \ - \ \frac{3}{z_m \ - \ z} \ , \qquad b \ = \ 0
\eeq
close to $z_m$, so that eq.~\eqref{V_schrod} implies that
\beq
\tilde{\mu} \ = \ 1 \ .
\eeq
\item On the other hand, for $\gamma > \gamma_c$ the preceding expressions imply that  for tensor perturbations
\beq
a \ = \ \frac{8 \,\gamma_c^2}{z\left(9 \,\gamma^2 \ - \ \gamma_c^2\right)} \ , \qquad b \ = \ 0
\eeq
close to the origin, so that, on account of eq.~\eqref{V_schrod},
\beq
\mu \ = \ \frac{9}{2} \ \frac{\gamma^2 \ - \ \gamma_c^2}{9\, \gamma^2 \ - \ \gamma_c^2} \ ,
\eeq
and clearly $0 \leq \mu \leq \frac{1}{2}$. On the other hand
\beq
a \ = \ - \ \frac{1}{z_m \ - \ z} \ , \qquad b \ = \ 0
\eeq
close to $z_m$, so that eq.~\eqref{V_schrod} implies that
\beq
\tilde{\mu} \ = \ 0 \ .
\eeq
For scalar perturbations
\beq
a \ = \ \frac{6}{z}\ \frac{3\,\gamma^2\ +\ \gamma_c^2}{9\, \gamma^2 \ - \ \gamma_c^2} \ , \qquad b \ = \ 0
\eeq
close to the origin, so that eq.~\eqref{V_schrod} implies that
\beq
\mu \ = \ \frac{9\,\gamma^2 \ + \ 7\,\gamma_c^2}{2\left(9\, \gamma^2 \ - \ \gamma_c^2\right)} \ ,
\eeq
and clearly $\frac{1}{2} \leq \mu \leq 1$, while
\beq
a \ = \ - \ \frac{3}{z_m \ - \ z} \ , \qquad b \ = \ 0
\eeq
close to $z_m$, so that eq.~\eqref{V_schrod} implies that
\beq
\tilde{\mu} \ = \ 1 \ .
\eeq
Table~\ref{tab:tab_munu} summarizes the values of $\mu$ and $\tilde{\mu}$ for tensor and scalar perturbations in the different ranges for $\gamma$.
\begin{table}[h!]
\centering
\begin{tabular}{||c || c | c || c | c||}
 \hline
Range & $\mu_T$ & $\tilde{\mu}_T$ & $\mu_S$ & $\tilde{\mu}_S$  \\ [0.5ex]
 \hline\hline
$\gamma \leq \gamma_c $ & $0$ &  $0$ & $1$ & $1$ \\ \hline
$\gamma > \gamma_c $ & $\frac{9}{2} \ \frac{\gamma^2 \ - \ \gamma_c^2}{9\, \gamma^2 \ - \ \gamma_c^2}$ &  $0$ & $\frac{9\,\gamma^2 \ + \ 7\,\gamma_c^2}{2\left(9\, \gamma^2 \ - \ \gamma_c^2\right)}$ & $1$  \\
 & \small{${\left(0 \leq \mu_T \leq \frac{1}{2}\right)}$} & & \small{$\left( \frac{1}{2} \leq \mu_S \leq {1}\right)$} &
\\
 [1ex]
 \hline
\end{tabular}
\caption{Values of $\mu$ and $\tilde{\mu}$ for the different ranges of $\gamma$ for tensor and scalar perturbations.}
\label{tab:tab_munu}
\end{table}
\item In addition, for the orientifold models of~\cite{as95,sugimoto}, with $\gamma=\gamma_c=\frac{3}{2}$, one finds
\bea
 \mu_V &=& \ \frac{3}{8} \ , \qquad\qquad\qquad \tilde{\mu}_V  \ = \  \frac{3}{8} \ , \nonumber \\
\mu_{RR} &=& \ \frac{5}{8}\, \qquad\qquad\qquad  \ \tilde{\mu}_{RR} \ = \ \frac{7}{8} \ ,
\eea
for vector and Ramond--Ramond perturbations,
while for the $SO(16) \times SO(16)$ heterotic model of~\cite{so1616}, with $\gamma=\frac{5}{2}$, one finds
\bea
\mu_V &=& \frac{1}{8} \ , \qquad\qquad\qquad \ \ \ \tilde{\mu}_V \ = \ \frac{3}{8} \ , \nonumber \\
  \mu_{NS} &=& \frac{1}{24} \ , \qquad\qquad\qquad \tilde{\mu}_{NS} \ = \ \frac{7}{8} \ ,
\eea
 for vector and NS-NS perturbations.
\end{itemize}

\section{\sc Self--Adjoint Schr\"odinger Systems} \label{sec:selfadjoint}

In~\cite{bms,ms22_1}, as well as in the previous section, we discussed how to define Hermitian Schr\"odinger--like settings for Bose and Fermi modes combining suitable choices of the independent variable with redefinitions of the fields. Here we would like to supplement those results describing how to attain self--adjoint extensions, referring to a large extent to the finite intervals that are central to the present setting. Self-adjointness requires proper choices of boundary conditions, and is instrumental to identify complete sets of modes. These are needed to expand arbitrary perturbations, and thus to make definite statements on the issue of stability that is central to the present work and to~\cite{bms,ms22_1,ms23_1}. Most of the basic setup is dealt with at length in the literature, and in particular in~\cite{math_literature_1} (see also the more recent treatments in~\cite{math_literature_2}), but here we shall reformulate it in a way that seems more transparent to us.

Our aim, in this section, is highlighting how proper choices of boundary conditions at the endpoints of an interval $[0,z_m]$ of the real axis can grant that a Schr\"odinger--like operator
\beq
H \ = \ - \ \partial_z^{\,2} \ + \ V(z) \label{schb1}
\eeq
be self-adjoint. To this end, let us first consider a potential $V(z)$ that is regular in the interval, together with a pair of wavefunctions $\psi$ and $\chi$ with the standard $L^2$ scalar product.

The condition granting that $H$ be Hermitian is the vanishing of the boundary contribution
\beq
\Big[ \psi^\star \, \partial_z \, \chi \ - \ \partial_z \,\psi^\star \ \chi \Big]_0^{z_m} \ = \ 0 \ , \label{b.2}
\eeq
for all $\psi$ and $\chi$ wavefunctions belonging to the domain ${\cal D}$ of $H$. This domain identifies, in general, sets of functions $\psi$ subject to proper boundary conditions that are in $L^2$, and such that $H\,\psi$ and $H\,\chi$ are also in $L^2$.

$H$ is Hermitian if, for any $\chi$ and $\psi$ in ${\cal D}$, eq.~\eqref{b.2} holds, but this condition does not suffice to guarantee the self-adjointness.
The operator $H$ is self--adjoint if, given any element $\chi$ in $L^2$ such that $H\,\chi$ is in $L^2$ and the condition~\eqref{b.2} holds for all $\psi$ belonging to ${\cal D}$, then $\chi$ must also belong to ${\cal D}$. Defining the two doublets
\beq
\underline{\psi} \ = \ \left( \begin{array}{c} \psi \\ z_m\ \partial_z \,\psi \end{array} \right) \ , \qquad\qquad \underline{\chi} \ = \ \left( \begin{array}{c} \chi \\ z_m\ \partial_z \,\chi \end{array} \right) \ , \label{psichi}
\eeq
the condition~\eqref{b.2} can be cast in the convenient form
\beq
\underline{\psi}^\dagger\left(z_m\right) \ \sigma_2\ \underline{\chi}\left(z_m\right) \ = \ \underline{\psi}^\dagger\left(0\right) \ \sigma_2\ \underline{\chi}\left(0\right) \ ,  \label{sigma2}
\eeq
with $\sigma_2$ the usual Pauli matrix.

The linear boundary conditions defining ${\cal D}$ for self--adjoint extensions of the Schr\"odinger operator~\eqref{schb1} can thus be described by $SU(1,1)$-valued, or equivalently by $SL(2,R)$-valued matrices $U$, endowed with an additional phase factor $\beta$, such that
\beq
U^\dagger \ \sigma_2 \ U \ = \ \sigma_2 \ , \label{b5}
\eeq
and the $\chi$ and $\psi$ wavefunctions satisfy
\beq
e^{i\beta}\,\underline{\psi}\left(z_m\right) \ = \ U \, \underline{\psi}\left(0\right) \ , \qquad e^{i\beta}\,\underline{\chi}\left(z_m\right) \ = \ U \, \underline{\chi}\left(0\right)  \ . \label{bcU}
\eeq
These boundary conditions depend on four real numbers parametrizing  $U(1,1)$ and relate, in general, the values of the wavefunction and its derivative at $z=0$ to those at $z=z_m$, rather than specifying them independently at the two ends.

A global characterization of $SL(2,R)$, whose group manifold can be identified with $AdS_3$, is given by
\beq
U\left(\rho,\theta_1,\theta_2\right) \ = \ \cosh\rho\left(\cos\theta_1 \, \underline{1} \,-\, i\,\sigma_2\,\sin\theta_1\right) \,+\, \sinh\rho\left(\sigma_3\, \cos\theta_2\ + \ \sigma_1\, \sin\theta_2 \right)
\ , \label{global_ads3_1}
\eeq
where $0 \leq \rho < \infty$, $0\leq \theta_{1}< 2\,\pi$, $0\leq \theta_{2}< 2\,\pi$, and the inverse is
\beq
U^{-1}\left(\rho,\theta_1\,\theta_2\right)\,=\,U\left(\rho,-\theta_1,\theta_2+\pi \right) \ . \label{bcUm}
\eeq
An important class of $U$ matrices emerges from the preceding expressions in the $\rho \to \infty$ limit, where
\beq
U \ \sim \ \frac{1}{2}\,e^{\,\rho} \, {\cal M}\left(\theta_1,\theta_2\right) \ ,  \label{Ulim}
\eeq
with
\beq
{\cal M}\left(\theta_1,\theta_2\right) \ = \ \left(\begin{array}{cc} \cos\theta_1+\cos\theta_2 & \sin\theta_2-\sin\theta_1 \\ \sin\theta_1+ \sin\theta_2 & \cos\theta_1-\cos\theta_2 \end{array} \right) \ .
\eeq
Making use of eqs.~\eqref{bcU}, \eqref{bcUm} and \eqref{Ulim}, one thus recovers the more familiar local boundary conditions, given independently at the two ends:
\beq
{\cal M}\left(\theta_1,\theta_2\right) \underline{\psi}\left(0\right) \ = \ 0 \ , \qquad {\cal M}\left(-\,\theta_1,\theta_2\,+\,\pi\right) \underline{\psi}\left(z_m\right) \ = \ 0 \ .
\eeq
Only two of these conditions are independent, since the ${\cal M}$ matrices have rank one and their product vanishes, so that one can work with
\bea
&& \left(\cos\theta_1\ + \ \cos\theta_2\right)\psi\left(0\right) \ + \ \left(\sin\theta_2 - \ \sin\theta_1\right)z_m \partial_z\,\psi\left(0\right)  \ = \ 0 \ , \nonumber \\
&& \left(\cos\theta_1\ - \ \cos\theta_2\right)\psi\left(z_m\right)  \ + \ \left(\sin\theta_1 \ - \ \sin\theta_2\right)z_m \,\partial_z\,\psi\left(z_m\right)  \ = \ 0 \ , \label{cond1}
\eea
or alternatively with
\bea
&& \left(\sin\theta_1\ + \ \sin\theta_2\right)\psi\left(0\right) \ + \ \left(\cos\theta_1\  - \ \cos\theta_2\right)z_m \partial_z\,\psi\left(0\right)  \ = \ 0 \ , \nonumber \\
&& \left(\sin\theta_1\ + \ \sin\theta_2\right)\psi\left(z_m\right)  \ - \ \left(\cos\theta_1 \ + \ \cos\theta_2\right)z_m \,\partial_z\,\psi\left(z_m\right)  \ = \ 0 \ . \label{cond2}
\eea
One can use either pair, unless one condition vanishes identically. Some special choices recover the familiar Dirichlet (D) and Neumann (N) conditions: $\theta_1=\theta_2= \frac{\pi}{2}$ yields DD conditions, $\theta_1=- \theta_2= \frac{\pi}{2}$ yields NN conditions, and finally $\theta_1= \theta_2= 0$ and $\theta_1= 0$, $\theta_2= \pi$ lead to DN and ND conditions.

Each choice of $U$ and the phase $\beta$ defines a self--adjoint extension of $H$, and the resulting spectrum depends, in general, on four real parameters.  There are some surprises, as we shall see shortly. For example, one or even two negative eigenvalues can emerge, even in the simple case $V=0$, with suitable choices of boundary conditions.

\subsection{\sc The Free Theory in an interval} \label{sec:free}

It is instructive to take a close look at the simple Hamiltonian
\beq
H \ = \ - \ \partial_z^2 \ ,
\eeq
in the interval $[0,z_m]$, for the general self--adjoint boundary conditions in eq.~\eqref{bcU}, so that the eigenvalue equation of interest is
\beq
H \, \psi(z) \ = \ k^2\, \psi(z) \ .
\eeq

As we have seen in eq.~\eqref{bcU}, general self--adjoint boundary conditions involve a phase $\beta$ and generic $SL(2,R)$ matrices $U$. These boundary conditions can be described via eq.~\eqref{global_ads3_1}, or alternatively via a vector $\vec{n}$, writing the $SL(2,R)$ matrix as
\beq
U\left(\vec{n}\right) \ = \ e^{\,n_1\,\sigma_1 \ + \ i\,n_2\,\sigma_2 \ + \ n_3\,\sigma_3} \ = \ e^{\vec{n}\,\cdot\, \vec{\sigma}} \ , \label{Un}
\eeq
with $\vec{n}=(n_1,i\,n_2,n_3)$ and real $n_i$, which can be more convenient in some cases.

The eigenvalue equation for the vector $\underline{\psi}$ of eq.~\eqref{psichi} can be recast into the first--order system
\beq
z_m\,\partial_z\, \underline{\psi} \ = \ \left[ \left(\begin{array}{cc} 0 & 1 \\ 0 & 0\end{array}\right) \ - \ \left(k\,z_m\right)^2 \left(\begin{array}{cc} 0 & 0 \\ 1 & 0\end{array}\right) \right] \underline{\psi} \ \equiv \ \left[ \sigma_+ \ - \ \left(k\,z_m\right)^2 \sigma_-\right] \underline{\psi} \ ,
\eeq
and in this notation the general solution reads
\beq
\underline{\psi}(z) \ = \ V(z) \, \underline{\psi}(0) \ , \label{Vmatrix}
\eeq
where
\beq
V(z) \ = \ \cos(k z) \, \underline{1} \ + \ \Big(\sigma_+\,-\,\left(k\,z_m\right)^2\,\sigma_-\Big) \frac{\sin(k z)}{k z_m}
\eeq
is also a special $SL(2,R)$ matrix. This is a general property for Hamiltonians of the form~\eqref{ham}, which reflects the constancy of the Wronskian. Consequently, one can write
\beq
V\left(z_m\right) \ = \ U\left(\vec{n}'\right) \ , \label{Vzm}
\eeq
where $\vec{n}'$ is a special $z$-dependent vector of the type introduced in eq.~\eqref{Un}, with components
\beq
{n}_1{}' \ = \ \frac{1}{2}\left(1 \ - \ \left(k\,z_m\right)^2 \right) \ , \qquad i\,{n}_2{}' \ = \ \frac{i}{2}\left(1 \ + \ \left(k\,z_m\right)^2 \right) \ , \qquad n_3{}'\ = \ 0   \ .
\eeq
As a result, the boundary condition~\eqref{bcU} translates into
\beq
U^{-1}\left(\vec{n}\right)\, U\left(\vec{n}{}'\right) \, \underline{\psi}(0) \ = \ e^{\,-\,i\,\beta}\ \underline{\psi}(0) \ , \label{eigenvone}
\eeq
and demands that one of the eigenvalues of the product of the two $U$ matrices, which is also an $SL(2,R)$ matrix, be equal to $e^{-\,i\,\beta}$.

When referring to eq.~\eqref{Un}, one must actually distinguish three classes of $SL(2,R)$ elements, according the value of $\vec{n} \cdot \vec{n}$.
\begin{itemize}
\item If $\vec{n} \cdot \vec{n} \ = \ 0$, the vector $\vec{n}$ depends on a real parameter $r$ and an angle $\theta$, according to
\beq
\vec{n} \ = \ \left(r\,\cos\theta,\pm\,i\,r,r\,\sin\theta \right) \ .
\eeq
In this case the $SL(2,R)$ matrix $U\left(\vec{n}\right)$ is
\beq
U\left(\vec{n}\right) \ = \ 1 \ + \ \vec{n}\cdot\vec{\sigma} \label{Unull}
\eeq
and has only the eigenvalue 1.
\item If $\vec{n} \cdot \vec{n} \ = \ \lambda^2 > 0$, the vector $\vec{n}$ depends on the two real parameters $\lambda>0$ and $\zeta$, and on an angle $\theta$, according to
\beq
\vec{n} \ = \ \lambda\left(\cos\theta\,\cosh\zeta, i\,\sinh\zeta,\sin\theta\,\cosh\zeta \right) \ .
\eeq
In this case the $SL(2,R)$ matrix $U\left(\vec{n}\right)$ is
\beq
U\left(\vec{n}\right) \ = \ \cosh\lambda \ + \ \frac{\sinh\lambda}{\lambda}\, \vec{n}\cdot\vec{\sigma} \ ,
\eeq
and its eigenvalues are $e^{\pm\lambda}$.
\item If $\vec{n} \cdot \vec{n} \ = \ - \,\lambda^2 < 0$, the vector $\vec{n}$ depends again on the two real parameters $\lambda>0$ and $\zeta$, and on an angle $\theta$, according to
\beq
\vec{n} \ = \ \lambda\left(\cos\theta\,\sinh\zeta, \pm\,i\,\cosh\zeta,\sin\theta\,\sinh\zeta \right) \ .
\eeq
In this case the $SL(2,R)$ matrix $U\left(\vec{n}\right)$ is
\beq
U\left(\vec{n}\right) \ = \ \cos\lambda \ + \ \frac{\sin\lambda}{\lambda}\, \vec{n}\cdot\vec{\sigma} \ ,
\eeq
and its eigenvalues are $e^{\pm\,i\,\lambda}$, so that $\lambda$ has effectively a limited range and behaves as an angle.
\end{itemize}

Note that the preceding considerations imply that the trace of $U$ uniquely determines its class, and thus its eigenvalues. In the three cases
\beq
\mathrm{Tr}\left(U\right)\left(\vec{n}\right) \ = \ \left\{2,2\,\cosh\lambda,2\,\cos\lambda \right\}
\ . \eeq
Consequently $U^{-1}\left(\vec{n}\right)\,U\left(\vec{n}'\right)$ is of the third type, and the condition~\eqref{eigenvone} is equivalent to
\beq
\mathrm{Tr}\left[U^{-1}\left(\vec{n}\right)\,U\left(\vec{n}'\right)\right] \ = \ 2\,\cos\beta \ , \label{eingevcond}
\eeq
which is the equation that determines the energy eigenvalues. Given any energy eigenvalue, eq.~\eqref{eigenvone} yields the corresponding boundary conditions.

We can now discuss in detail the content of eq.~\eqref{eingevcond}, referring to eq.~\eqref{kmu}
and using for $U\left(\vec{n}\right)$ the global parametrization of eq.~\eqref{global_ads3_1}.

\subsubsection{\sc Negative--Energy Solutions} \label{app:negfree}

For negative--energy states,  so that
\beq
\mu^2 \ = \ - \left(k z_m\right)^2  \label{kmu}
\eeq
is positive, eq.~\eqref{eingevcond} takes the form
\beq
\cosh\rho\left[\cos\theta_1 \cosh\mu \,{ - }\, \left(1-\mu^2\right) \frac{\sinh\mu}{2\,\mu}\,\sin\theta_1\right] \, { - } \, \left(1 + \mu^2\right)\frac{\sinh\mu}{2\,\mu}\,\sinh\rho\, \sin\theta_2 = \cos\beta \ , \label{neg_en_gen}
\eeq
with real values of $\mu$. The features of these negative--energy states depend on the values of a number of parameters, and by tuning them one can make their energy approach zero.
\begin{figure}[ht]
\centering
\includegraphics[width=50mm]{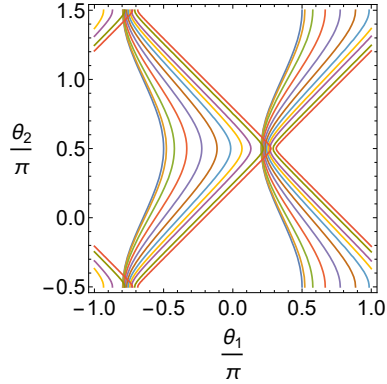}
\caption{\small Curves of constant tachyonic mass for the large--$\rho$ eigenvalue equation~\eqref{neg_en_gen_rhoinf}. The filled regions identify boundary conditions leading to instabilities.}
\label{fig:tachyon_free}
\end{figure}

In the $\mu\to 0$ limit, the eigenvalue equation simplifies considerably and becomes
\beq
\cosh\rho\left[\cos\theta_1 \,-\, \frac{1}{2}\,\sin\theta_1\right] \, - \, \frac{1}{2}\,\sinh\rho\, \sin\theta_2 = \cos\beta \ . \label{neg_simple}
\eeq
Letting
\beq
\sin\gamma \ = \ \frac{2}{\sqrt{5}} \ , \qquad \cos\gamma \ = \ \frac{1}{\sqrt{5}} \ , \label{zeromode}
\eeq
the simpler eigenvalue equation~\eqref{neg_simple} be cast in the more compact form
\beq
\sqrt{5}\,\cosh\rho\,\sin\left(\theta_1 \,-\, \gamma\right)  \, + \, \sinh\rho\, \sin\theta_2 \,= \,-\,2\,\cos\beta \ . \label{neg_compact}
\eeq
For example, there are two solutions of eq.~\eqref{neg_simple}, $\rho \simeq 0.69$ or $4.47$, if $\sin\left(\theta_1-\gamma\right)\simeq 0.3$, $\sin\theta_2\simeq - 0.65$, $\cos\beta \simeq -\,0.17$. There is always one solution if $\sin\left(\theta_1-\gamma\right)=0$, while if $\sin\theta_2=0$ there is one solution if $\frac{2\cos \beta}{\sqrt{5} \sin \left(\theta_1-\gamma\right)} \leq -\,1$ and there are no solutions otherwise. These zero--mass solutions are generically limits of tachyonic modes.
There can be tachyons with arbitrarily large $\mu^2$. A simple example obtains taking $ \cos\beta=\cos\theta_1=0$ and $\sin\theta_1=\sin\theta_2=1$, so that eq.~\eqref{neg_en_gen} reduces to $\mu^2 = e^{2\rho}$. There are then states whose negative energies span the whole range $(-\infty,0]$, depending on the value of $\rho$.

For large values of $\rho$ the eigenvalue equation~\eqref{neg_en_gen} simplifies and reduces to
\beq
\cos\theta_1 \cosh\mu \,{ - }\, \left(1-\mu^2\right) \frac{\sinh\mu}{2\,\mu}\,\sin\theta_1 \, { - } \, \left(1 + \mu^2\right)\frac{\sinh\mu}{2\,\mu}\,\sin\theta_2 = 0 \ , \label{neg_en_gen_rhoinf}
\eeq
and tachyonic modes are present in the filled region of fig.~\ref{fig:tachyon_free}, where the different curves correspond to different values of $\mu$. As $\mu$ increases, the tachyonic modes quickly pile up along the curve
\beq
\sin\theta_1 \ = \ \sin\theta_2 \ ,
\eeq
which corresponds to the straight lines in fig.~\ref{fig:tachyon_free}.

\subsubsection{\sc Positive--Energy Solutions and Exactly Solvable Cases} \label{sec:free_positive}

\begin{figure}[ht]
\centering
\begin{tabular}{cc}
\includegraphics[width=50mm]{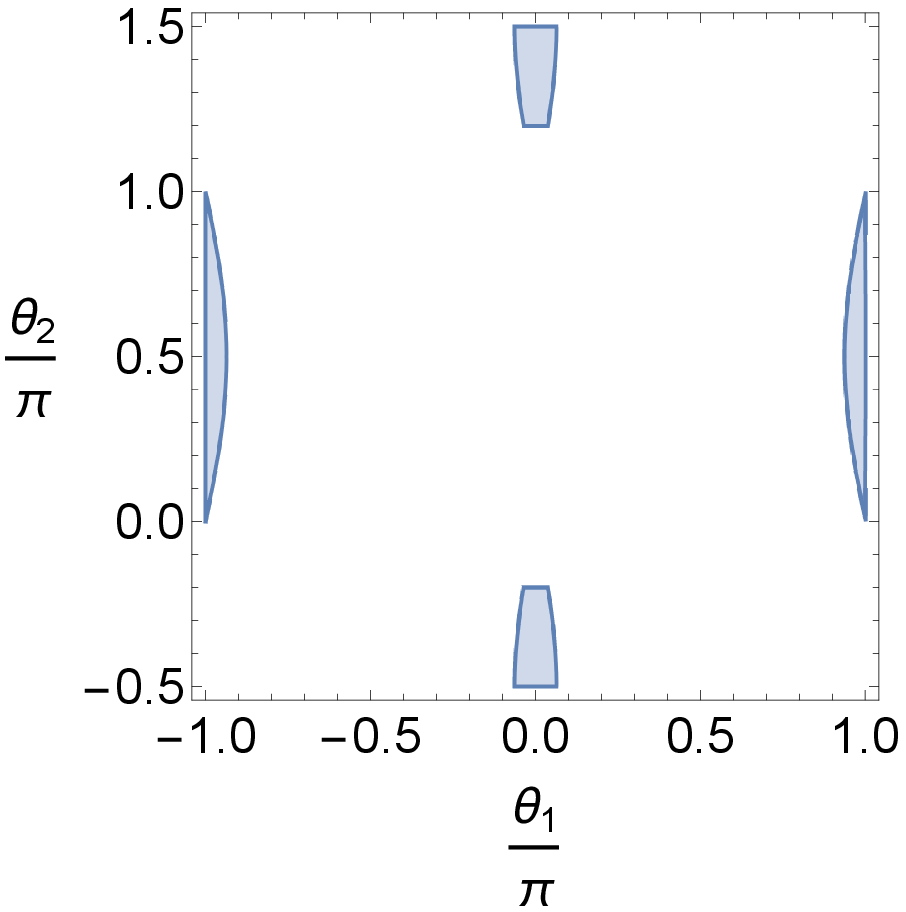} \qquad \qquad &
\includegraphics[width=50mm]{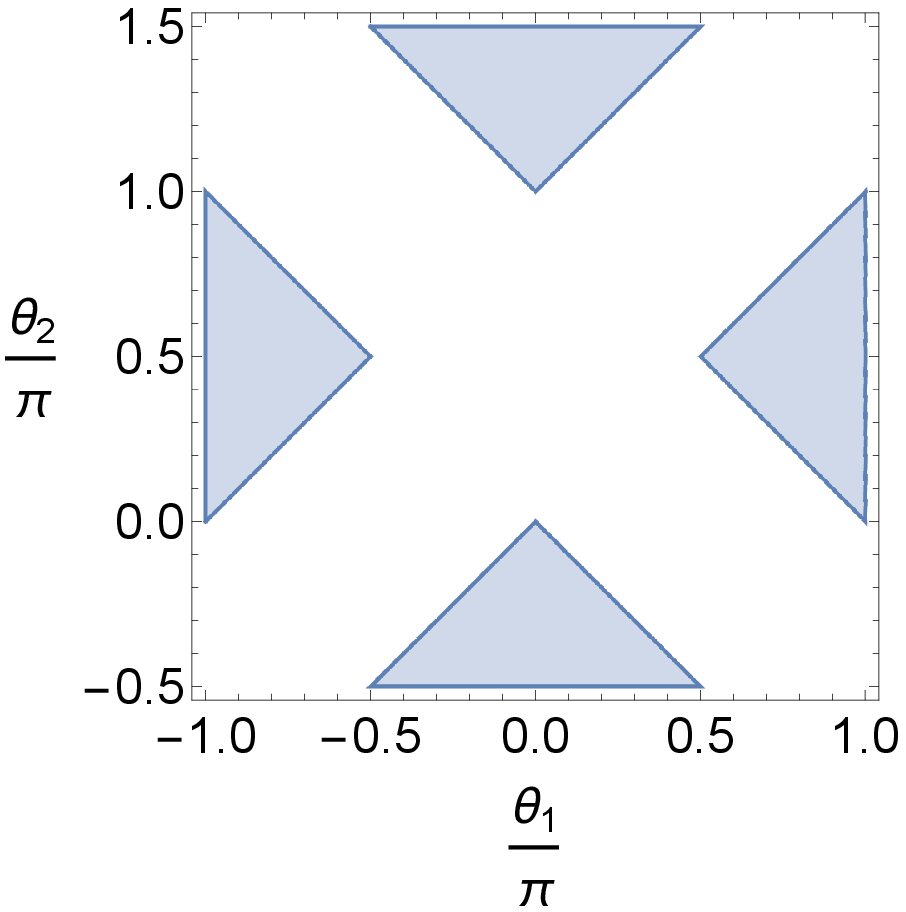} \\
\end{tabular}
\caption{\small The shaded areas are the regions of parameter space where positivity holds, for $\rho=0.2$ (left panel), and for large $\rho$ (right panel).}
\label{fig:positivity_free}
\end{figure}

It is now interesting to take a closer look at the choices of boundary conditions that yield positive spectra for $k^2$. To this end, let us note that, after a partial integration, for a normalized $\psi$,
\beq
k^2 \ = \ \int dz \left| \partial_z\,\psi\right|^2 \ - \ \left. \psi^\star\,\partial_z\,\psi\right|_{z=z_m} \ + \ \left. \psi^\star\,\partial_z\,\psi\right|_{z=0} \ .
\eeq
The positivity of $k^2$ is thus guaranteed if the boundary contribution is  positive. In terms of the two--component vector $\underline{\psi}$, this condition takes the form
\beq
- \ \left.\underline{\psi}^\dagger \left(\sigma_1 \,+\, i\,\sigma_2\right) \underline{\psi}\right|_{z=z_m} \ + \ \left.\underline{\psi}^\dagger \left(\sigma_1 \,+\, i\,\sigma_2\right) \underline{\psi}\right|_{z=0} \ \geq \ 0 \ .
\eeq
Making use of the self--adjoint boundary conditions~\eqref{bcU} and of eq.~\eqref{b5}, this reduces to requiring that
\beq
\left. \underline{\psi}^\dagger\left( \sigma_1 \ - \ U^\dagger\,\sigma_1\, U\right)\underline{\psi}\right|_{z=0} \ \geq \ 0  \ . \label{pos_psiund}
\eeq
The inequality is guaranteed to hold if the trace and the determinant of the symmetric matrix
\beq
{\cal S} \ = \ \sigma_1 \ - \ U^\dagger\,\sigma_1\, U
\eeq
are both non-negative. In terms of the global parametrization~\eqref{global_ads3_1}, these conditions translate into
\bea
&& \sin\left(\theta_1\ + \ \theta_2\right) \ \leq \ 0 \ , \nonumber \\
&&  \sin^2 \theta_1 \ - \ \tanh^2 \rho \, \sin^2 \theta_2  \ \leq \ 0 \ .  \label{pos_sl2r}
\eea
For $\rho \simeq 0$ the allowed region reduces to $\theta_1= 0,\pi$ and ${\pi} \,\leq \,\theta_1\,+\,\theta_2 \,\leq \, {2 \pi}$ mod $2\pi$, while for positive values of $\rho$ it rapidly approaches the profile displayed in the right panel of fig.~\ref{fig:positivity_free}. Note that the constant--$\mu$ curves of fig.~\ref{fig:tachyon_free} lie outside the shaded regions in fig.~\ref{fig:positivity_free}.

There are infinite spectra of positive--energy solutions, which can be investigated letting $\mu = i \,\nu$ with real $\nu$ in eq.~\eqref{neg_en_gen}, thus turning it into
\beq
\cosh\rho\left[\cos\theta_1 \cos\nu \,-\, \left(1+\nu^2\right) \frac{\sin\nu}{2\,\nu}\,\sin\theta_1\right] \, - \, \left(1 - \nu^2\right)\frac{\sin\nu}{2\,\nu}\,\sinh\rho\, \sin\theta_2 = \cos\beta \ . \label{b.75}
\eeq
The choices $\left(U,\beta\right)$ and $\left(-\,U,\beta\,+\,\pi \right)$ identify identical elements of $U(1,1)$, so that the independent parameters actually correspond to $SL(2,R) \times U(1)/Z_2$, for both positive--energy and negative--energy solutions.

In some cases, which we can now discuss, this eigenvalue equation can be solved relatively simply.
\begin{itemize}
\item The simplest case corresponds to
\beq
U \ = \ 1 \ , \qquad \beta \ = \ 0 \ ,
\eeq
and is recovered for $\rho=0$, $\theta_1=0$. The eigenvalue equation~\eqref{b.75} then reduces to
\beq
\cos \nu \ = \ 1 \ ,
\eeq
and consequently $\nu = 2\,\ell\,\pi$, with $\ell$ an arbitrary integer. Moreover, since $U\left(\vec{n}'\right)$ is the identity, $\underline{\psi}(0)$ is unconstrained and there are two independent solutions of this type for any choice of $\ell$. The end result is indistinguishable from the spectrum corresponding to a circle of radius $\frac{z_m}{2 \pi}$.
\item If  $U = 1$
but $\beta \neq 0$, there are two independent families of solutions, with
\beq
\nu_\pm = \pm \,\beta \ + \ 2\,\ell\,\pi \ ,
\eeq
and the preceding degeneracy is lifted, since $U\left(\vec{n}'\right)$ is no more the identity. The end result is indistinguishable from the spectrum corresponding to a circle of radius $\frac{z_m}{2\,\pi}$ with twisted periodicity.
\item If $\theta_1=0$ and $\sin\theta_2$ also vanishes but $\rho \neq 0$,
\beq
U\left(\rho,\theta_1,\theta_2\right) \ = \ \left(\begin{array}{cc} e^{\pm\,\rho} & 0 \\ 0 & e^{\mp \,\rho} \end{array}\right)
\ , \label{global_ads3_2}
\eeq
and the eigenvalue equation is solved by
\beq
\nu \ = \ \pm \ \arccos\left(\frac{\cos\beta}{\cosh\rho}\right) \ + \ 2\,\ell\,\pi \ . \label{2lpi}
\eeq
\item If $\theta_1=\pi$ and $\sin\theta_2$ also vanishes,
\beq
U\left(\rho,\theta_1,\theta_2\right) \ = \ - \ \left(\begin{array}{cc} e^{\mp\,\rho} & 0 \\ 0 & e^{\pm \,\rho} \end{array}\right)
\ , \label{global_ads3_3}
\eeq
and the eigenvalue equation is solved by
\beq
\nu \ = \ \pm \ \arccos\left(\frac{\cos\beta}{\cosh\rho}\right) \ + \ \left(2\,\ell\,+\,1\right)\pi \ . \label{2lpi+pi}
\eeq
\end{itemize}

\subsubsection{\sc Large--$\nu$ (or Semi--Classical) Behavior}

There are some simplifications in eq.~\eqref{b.75} for large values of $\nu$. One can distinguish two cases.
\begin{itemize}
    \item The terms proportional to $\nu\, \sin\nu$ are absent in eq.~\eqref{b.75} if
\beq
\cosh\rho\,\sin\theta_1 \ - \ \sinh\rho\,\sin\theta_2 \ = \ 0 \ , \label{b.78}
\eeq
or equivalently if $U(\vec{n})$ is lower triangular.
This is only possible if
\beq
\left|\sin\theta_1\right| \ < \ \left|\sin\theta_2\right| \ ,
\eeq
so that $\cos\theta_1$ cannot vanish, and then eq.~\eqref{b.75} reduces to
\beq
\cosh\rho\left[\cos\theta_1 \cos\nu \,-\, \frac{\sin\nu}{\nu}\,\sin\theta_1\right] \  = \ \cos\beta \ , \label{b.75n}
\eeq
There is an infinite number of allowed frequencies, whose behavior
for large values of $\nu$ is dominated by the first term, while the second determines the leading correction. Consequently
\beq
\nu_\ell^\pm \ = \ \pm \ \arccos\left(\frac{\cos\beta}{\cos\theta_1\,\cosh\rho}\right) \ + \ 2\,\ell\,\pi \ \equiv \ \nu_0^\pm \ + \ 2\,\ell\,\pi \ , \label{b.81}
\eeq
or, taking the leading correction into account,
\beq
\nu^\pm \ \simeq \ \nu_0^\pm \ + \ 2\,\ell\,\pi \ + \ \frac{\tan\theta_1 }{\nu_\ell^\pm}\ . \label{b.48}
\eeq

\item On the other hand, if eq.~\eqref{b.78} does not hold, which is the generic case, the terms proportional to $\nu\, \sin\nu$ must vanish to leading order, and the large eigenvalues of eq.~\eqref{b.75} approach
\beq
\nu \ \simeq \ \ell\,\pi \ - \ \frac{2}{\ell\,\pi} \ \frac{\cos\beta \ - \ (-1)^\ell \cosh\rho}{\cosh\rho\,\sin\theta_1 \ - \ \sinh\rho\,\sin\theta_2} \ , \label{b.49}
\eeq
with $\ell$ a large positive integer.
\end{itemize}
There are thus two classes of leading behavior for large $\nu$, which correspond asymptotically to the same number of modes.

\subsubsection{\sc Large--$\rho$ Behavior and Open Boundary Conditions} \label{sec:large rho}

As we have explained, the large--$\rho$ limit yields independent boundary conditions at the two ends of the interval that generalize the standard Neumann and Dirichlet choices.

After eq.~\eqref{Ulim} we identified four special cases:
\begin{itemize}
\item Dirichlet--Dirichlet boundary conditions, which obtain for $\theta_1=\theta_2=\frac{\pi}{2}$. No negative energy solutions emerge from eq.~\eqref{neg_en_gen}, while eq.~\eqref{b.75} reduces to
\beq
\frac{\sin\nu}{\nu} \ = \ 0 \ ,
\eeq
and thus yields the familiar positive--energy spectrum, with $\nu_\ell\,=\,\ell \, \pi$ and $\ell=1,2,\ldots$.
\item Neumann--Neumann boundary conditions, which obtain for $\theta_1=- \theta_2=\frac{\pi}{2}$. There are again no negative--energy solutions, while eq.~\eqref{b.75} reduces to
\beq
\nu\,{\sin\nu} \ = \ 0 \ ,
\eeq
so that now $\nu_\ell\,=\,\ell \, \pi$ and $\ell=0,1,2,\ldots$, which now include a zero mode.
\item Neumann--Dirichlet and Dirichlet--Neumann boundary conditions, which obtain for $\theta_1=0$ and $\theta_2=0,\pi$. The eigenvalue equation~\eqref{b.75} is identical in the two cases, and there are again no negative-energy solutions, while now
\beq
{\cos\nu} \ = \ 0 \ ,
\eeq
so that $\nu_\ell\,=\,\left(\ell\,+\,\frac{1}{2}\right) \pi$ and $\ell=0,1,2,\ldots$.
\end{itemize}

Negative--energy solutions can still emerge in the $\rho \to \infty$ limit. This can be foreseen from the limiting form of eq.~\eqref{neg_compact} as $\rho \to \infty$,
\beq
\sqrt{5}\,\sin\left(\theta_1\,-\,\gamma\right) \ + \ \sin\theta_2 \ = \ 0 \ ,
 \eeq
which can admit at most two solutions, provided $\left|\sin\left(\theta_1\,-\,\gamma\right) \right| \leq \frac{1}{\sqrt{5}}$.
An interesting example of their occurrence is found letting $\theta_1=\frac{\pi}{2}$, which is relevant to both the Neumann--Neumann and Dirichlet--Dirichlet cases. In the large $\rho$ limit the eigenvalue equation~\eqref{neg_en_gen} is then exactly solved by
\beq
\mu^2 \ = \ \frac{1\ + \ \sin\theta_2}{1\ - \ \sin\theta_2} \ .
\eeq
These values of $\mu^2$ range from 0, the massless Neumann--Neumann mode, to $\infty$ where the Dirichlet--Dirichlet case is approached. The behavior for $\theta_2= \frac{\pi}{2} \pm \epsilon$, with $0<\epsilon \ll 1$ is peculiar, with two solutions associated to the two signs that are localized at the ends of the interval and disappear as $\epsilon \to 0$, but whose squares approach $\delta$ functions there.

The positive energy solutions of this type are determined in general by the $\rho \to \infty$ limit of eq.~\eqref{b.75},
\beq
\left[\cos\theta_1 \cos\nu \,-\, \left(1+\nu^2\right) \frac{\sin\nu}{2\,\nu}\,\sin\theta_1\right] \, - \, \left(1 - \nu^2\right)\frac{\sin\nu}{2\,\nu}\, \sin\theta_2 \ = \ 0 \ , \label{b.752}
\eeq
so that $\beta$ becomes irrelevant for these modes. There are two special cases in which the equation simplifies greatly. If $\sin\theta_1=\sin\theta_2$ it reduces to
\beq
\frac{\tan \nu}{\nu} \ = \ \cot\theta_1 \ ,
\eeq
while if $\sin\theta_1=- \sin\theta_2$ it reduces to
\beq
\nu\,{\tan \nu}\ = \ \cot\theta_1 \ ,
\eeq
and in both cases there are clearly infinitely many solutions, which approach $\left(\ell+\frac{1}{2}\right)\pi$ and $\ell\,\pi$, respectively. When $\sin\theta_1\neq \sin\theta_2$, the large--$\nu$ behavior is determined by the limiting form of eq.~\eqref{b.49}, and
\beq
\nu \ \simeq \ \ell\,\pi \ + \ \frac{2\,(-1)^\ell}{\ell\,\pi} \ \frac{1}{\sin\theta_1 \ - \ \sin\theta_2} \ . \label{b.492}
\eeq

\subsection{\sc Singular Potentials}

In all cases considered in Section~\ref{sec:sing_pot}, and more generally for the vacua in~\cite{dm_vacuum, ms21_1,ms21_2}, the potential $V(z)$ is singular at the ends of the interval, which we can denote by $z=0$ and $z=z_m$ when the range is finite, where it has double poles. The leading behavior depends, in general, on two real parameters $\mu$ and $\tilde{\mu}$, defined in eq.~\eqref{limiting}, whose values are summarized in Table~\ref{tab:tab_munu}.

The condition that $H\,\psi$ be in $L^2$ constrains in general the choice of wavefunctions~\cite{math_literature_1,math_literature_2}. Confining our attention to real and non--negative values for $\mu^2$ and $\tilde{\mu}^2$, which characterize all cases of interest for our problem, the asymptotic behaviors at the two ends of the $z$ interval are
\bea
\psi &\sim \  C_1\ \left(\frac{z}{z_m}\right)^{\frac{1}{2} \ + \ \mu} \ + \  C_2\ \left(\frac{z}{z_m}\right)^{\frac{1}{2} \ - \ \mu} \qquad\qquad\qquad\qquad &\mathrm{if} \qquad 0 \ < \ \mu \ < \ 1  \ ; \nonumber \\
\psi &\sim \  C_1\ \left(\frac{z}{z_m}\right)^{\frac{1}{2}} \ + \  C_2\ \left(\frac{z}{z_m}\right)^{\frac{1}{2}}\,\log \,\left(\frac{z}{z_m}\right)  \quad\qquad\qquad\qquad &\mathrm{if} \qquad \mu \ = \ 0 \ ; \nonumber \\
\psi &\sim \ C_3\ \left(1\,-\,\frac{z}{z_m}\right)^{\frac{1}{2} \ + \ \tilde{\mu}} \ + \  C_4\ \left(1\,-\,\frac{z}{z_m}\right)^{\frac{1}{2} \ - \ \tilde{\mu}}  \qquad\qquad  & \mathrm{if} \qquad 0 \ < \ \tilde{\mu} \ < \ 1 \ ; \nonumber \\
\psi &\sim \ C_3\ \left(1\,-\,\frac{z}{z_m}\right)^{\frac{1}{2}} \ + \  C_4\ \left(1\,-\,\frac{z}{z_m}\right)^{\frac{1}{2}}\,\log \left(1\,-\,\frac{z}{z_m}\right)  \quad &\mathrm{if} \qquad \tilde{\mu} \ = \ 0 \,.
\label{b.11}
\eea

For brevity, we can concentrate on different ranges for $\mu$, since identical considerations hold for $\tilde{\mu}$. If $\mu \geq 1$ only the solution behaving as $z^{\frac{1}{2} \ + \ \mu}$ is in $L^2$, so that there is a unique choice of limiting behavior.

Imaginary values of $\mu$ and $\tilde{\mu}$ are not encountered in the vacua of~\cite{dm_vacuum, ms21_1,ms21_2}, but have been examined in the literature~\cite{math_literature_1,math_literature_2} and lead the ``fall to the center'', with the inevitable emergence of infinitely many unstable modes. Leaving aside this pathological range, which does not concern the vacua of interest to us, there are still, in principle, four types of behavior, which we can now analyze separately.
\subsubsection{\sc $0 \leq \mu < 1$ and $0 \leq \tilde{\mu} < 1$}
This first case presents itself when $0 \leq \mu < 1$ and $0 \leq \tilde{\mu} < 1$, and one must still distinguish three sub-cases.
\begin{enumerate}
\item If $\mu=\tilde{\mu}=0$ the boundary term~\eqref{b.2} originating from two functions $\psi$ and $\chi$ as in eqs.~\eqref{b.11} is proportional to
\beq
C_1^\star\, D_2 \ - \ C_2^\star\, D_1 \ - \ C_4^\star\, D_3 \ + \ C_3^\star\, D_4 \ , \label{self-adj1}
\eeq
where the $C$'s correspond to the $\psi$'s and the $D$'s correspond to the $\chi$'s. The reader may recognize that this and the following expressions characterizing the self--adjoint boundary conditions rest
 on the Wronskian for a Hamiltonian $H$ as in eq.~\eqref{ham}, which is guaranteed to be non-singular even in the presence of a singular potential.
\item If $\mu \neq 0$ but $\tilde{\mu}=0$ the preceding boundary term becomes
\beq
 - \ 2\mu\left(C_1^\star\, D_2 \ - \ C_2^\star\, D_1\right) \ - \ C_4^\star\, D_3 \ + \ C_3^\star\, D_4 \ . \label{self-adj2}
\eeq
\item Similarly, if $\mu=0$ and $\tilde{\mu} \neq 0$ it becomes
\beq
C_1^\star\, D_2 \ - \ C_2^\star\, D_1 \  -  \ 2\tilde{\mu}\left(-\ C_4^\star\, D_3 \ + \ C_3^\star\, D_4\right) \ . \label{self-adj3}
\eeq
\item Finally, if $\mu \neq 0$ and $\tilde{\mu} \neq 0$ the boundary term becomes
\beq
2\,\mu\left(C_1^\star\, D_2 \ - \ C_2^\star\, D_1\right) \ + \ 2\tilde{\mu}\left(- \ C_4^\star\, D_3 \ + \ C_3^\star\, D_4\right) \ .\label{self-adj4}
\eeq
\end{enumerate}

Note that the $\mu \to 0$ and $\tilde{\mu} \to 0$ limits are singular, so that the last expression does not approach eq.~\eqref{self-adj1}. However, all these cases afford a unified formulation, if one lets
\beq
\underline{C}(0) \ = \ \sqrt{2\,\mu} \left(\begin{array}{c} C_1 \\ C_2  \end{array} \right) \quad \mathrm{if}\ \mu \neq 0\  , \qquad \underline{C}(0) \ = \ \left(\begin{array}{c} C_{2} \\ C_{1}  \end{array} \right) \quad \mathrm{if}\ \mu = 0 \ ,
\eeq
and
\beq
\underline{C}\left(z_m\right) \ = \ \sqrt{2\,\tilde{\mu}} \left(\begin{array}{c} C_4 \\ C_3  \end{array} \right) \quad \mathrm{if}\ \tilde{\mu} \neq 0\  , \qquad \underline{C}\left(z_m\right) \ = \ \left(\begin{array}{c} C_{3} \\ C_{4}  \end{array} \right) \quad \mathrm{if}\ \tilde{\mu} = 0 \ ,
\eeq
so that the self--adjointness condition becomes in all these cases
\beq
\underline{C}^\dagger(0)\ \sigma_2 \ \underline{D}(0) \ = \ \underline{C}^\dagger\left(z_m\right)\ \sigma_2 \ \underline{D}\left(z_m\right) \ , \label{P_constraint}
\eeq
which is similar to eq.~\eqref{sigma2}. All the preceding considerations thus apply, up to replacement of $\underline{\psi}(0)$ with the column vector $\underline{C}(0)$, and so on, so that one is led to boundary conditions like those in eq.~\eqref{bcU},
\beq
e^{i \beta}\, \underline{C} \left(z_m\right) \ = \ U \,\underline{C}\left(0\right) \ , \label{selfadjbeta}
\eeq
where $U$ is a generic $SL(2,R)$ matrix, which can be parametrized as in eq.~\eqref{global_ads3_1}, and $\beta$ is a phase. The large--$\rho$ limit yields again independent boundary conditions at the ends with two parameters $\theta_1$ and $\theta_2$. These are the counterparts of eqs.~\eqref{cond1} and \eqref{cond2}, and like them can be cast in the form
\bea
&& \cos\left(\frac{\theta_1-\theta_2}{2}\right) C_1 \ - \ \sin\left(\frac{\theta_1-\theta_2}{2}\right)C_2  \ = \ 0 \ , \nonumber \\
&&  \sin\left(\frac{\theta_1+\theta_2}{2}\right) C_4 \ - \ \cos\left(\frac{\theta_1+\theta_2}{2}\right)C_3  \ = \ 0    \ . \label{cond_sing1n}
\eea

The options corresponding to the DD, NN, ND and DN cases are $(C_1,C_4)=(0,0)$, $(C_2,C_3)=(0,0)$, $(C_2,C_4)=(0,0)$ and $(C_1,C_3)=(0,0)$.

\subsubsection{\sc $\mu \geq 1$ and $0 \leq \tilde{\mu} < 1$, or $0 \leq \mu < 1$ and $\tilde{\mu} \geq 1$}
These cases only differ by the interchange of the two ends, and present themselves when $\mu > 1$ and $0 \leq \tilde{\mu} \leq 1$, or when $0 \leq \mu \leq 1$ and $\tilde{\mu} > 1$.
Now the limiting behavior at one end is fixed and the corresponding boundary term~\eqref{b.2} vanishes. Eq.~\eqref{b.2} then reduces to the contribution at the other end, where two types of asymptotic behavior are still allowed. A similar limiting form obtains for $\chi$, with $C_{1,2}$ simply replaced by another pair of constants $D_{1,2}$. The result is thus proportional to
\beq
C_1^\star \, D_2 \ - \ C_2^\star \, D_1\ = \ \underline{C}^\dagger \ i\,\sigma_2 \ \underline{D} \ , \label{b.12}
\eeq
where, for instance
\beq
\underline{D} \ = \ \left( \begin{array}{c} D_1 \\ D_2 \end{array} \right) \ .
\eeq

This boundary term is eliminated if $\underline{C}$ and $\underline{D}$ are such that
\beq
\underline{C} \ = \ \Lambda \, \underline{C} \ , \qquad \underline{D} \ = \ \Lambda \, \underline{D} \ , \label{b.14}
\eeq
where $\Lambda$ is Hermitian, satisfies $\Lambda^2=1$ and anticommutes with $\sigma_2$, so that
\beq
\Lambda \ = \ \cos \alpha \ \sigma_3 \ + \ \sin\alpha \ \sigma_1 \ ,
\eeq
with $\alpha$ an arbitrary parameter. Each choice of $\alpha$ leads to a Hermitian Schr\"odinger--like operator, with
\beq
C_2 \ = \ \tan\left(\frac{\alpha}{2}\right) C_1 \ .
\eeq
In addition, if $\underline{C}$ verifies eq.~\eqref{b.14} and eq.~\eqref{b.12} is to vanish for all such choices, $\underline{D}$ must also satisfy eq.~\eqref{b.14}. In conclusion, one is thus led to \emph{a family of self--adjoint extensions} characterized by the angle $\alpha$.
\subsubsection{\sc $\mu \geq 1$ and $\tilde{\mu} \geq 1$ }
In this case the limiting behavior is fixed at
both ends, with exponents $\frac{1}{2}+\mu$ and $\frac{1}{2}+\tilde{\mu}$, while the boundary term vanishes identically.

Referring to Table~\ref{tab:tab_munu}, tensor perturbations fall into case 1 for all values of $\gamma$, while scalar perturbations fall into cases 3 or 2 depending on whether $\gamma \leq \gamma_c$ or $\gamma > \gamma_c$. In addition, vector and RR perturbations in the ten--dimensional orientifolds, and also vector and NS perturbations in the $SO(16)\times SO(16)$ model, belong to case 1.

\subsubsection{\sc A Family of Exactly Solvable Cases} \label{sec:exact}

As summarized in Table~\ref{tab:tab_munu}, within the range $\gamma \leq \gamma_c$ that is relevant for the orientifold models, the limiting behavior of tensor and scalar perturbations near the ends of the interval is as in eq.~\eqref{limiting}, with $\mu=\tilde{\mu}=0$ for the former and $\mu=\tilde{\mu}=1$ for the latter. It is thus interesting to examine the two classes of exactly solvable potentials
\beq
V_\pm \ = \ \left(\frac{\pi}{z_m}\right)^2 \left[ \frac{\left(\mu^2 \ - \ \frac{1}{4}\right)}{\sin^2\left( \frac{\pi\,z}{z_m}\right)} \ - \  \beta_\pm^2\right] \label{pot_sin}
\eeq
that have double poles at the ends of the interval with $\mu=\tilde{\mu}$, as in the cases of interest. The two constants
\beq
\beta_\pm \ = \ \frac{1}{2} \pm \ \mu  \label{betapm}
\eeq
are introduced so that these Hamiltonian are also expressible as products of first--order operators, according to
\beq
H_\pm \ \psi \ = \  {\cal A}_\pm \, {\cal A}^\dagger_\pm \,\psi \ = \ \left(\frac{ \pi\,m}{z_m}\right)^2 \  \psi \ . \label{b.28}
\eeq
Here $m$ is a dimensionless quantity proportional to the mass eigenvalues of $H_\pm$, and
\beq
{\cal A}_\pm \ = \ \partial_z \ + \ \frac{\pi}{z_m}\, {\beta_\pm}\,\cot\left(\frac{\pi\,z}{z_m}\right) \ , \qquad {\cal A}_\pm^\dagger \ = \ - \  \partial_z \ + \  \frac{\pi}{z_m}\,{\beta_\pm}\,\cot\left(\frac{\pi\,z}{z_m}\right) \ .
\eeq
Clearly,
\beq
H_+ \ - \ H_- \ = \ - \ 2\,\mu \left(\frac{ \pi}{z_m}\right)^2 \ ,  \label{delta_H}
\eeq
so that a zero mode of $H_+$ is also a massive mode of $H_-$, while a zero mode of $H_-$ is also a tachyonic mode of $H_+$.

For $0 < \mu<1$ each of the two Hamiltonians $H_\pm$ has a pair of normalizable zero modes, one of which solves ${\cal A}_\pm^\dagger \, \psi_\pm^{(0)} = 0$ in the two cases and reads
\beq
\psi_\pm^{(0)} \ = \ C \left[\sin\left(\frac{\pi\,z}{z_m}\right) \right]^{\beta_\pm} \ . \label{zeromodepsi}
\eeq
These two solutions correspond to special choices of the $C_i$ coefficients, which enter the self--adjoint extensions in eqs.~\eqref{b.11}. For the former, $\psi_+^{(0)}$, $C_2=C_4=0$, while for the latter, $\psi_-^{(0)}$, $C_1=C_3=0$. These boundary conditions are special choices corresponding to the $\rho \to \infty$ limit, and in the notation of eqs.~\eqref{cond_sing1n} they can be recovered for $(\theta_1,\theta_2)=(\pi,0)$ and for $(\theta_1,\theta_2)=(0,0)$.
There are two other zero-mode solutions, which can be constructed with the Wronskian method.
Moreover, on account of eq.~\eqref{delta_H}
\beq
H_+ \,\psi_-^{(0)} \ = \ - \ 2\,\mu \left(\frac{ \pi}{z_m}\right)^2\, \psi_-^{(0)} \ , \qquad
H_- \,\psi_+^{(0)} \ = \ 2\,\mu \left(\frac{ \pi}{z_m}\right)^2\, \psi_+^{(0)} \ ,
\eeq
so that $\psi_-^{(0)}$ is also a tachyonic mode of $H_+$, while $\psi_+^{(0)}$ is also a massive mode of $H_-$. Consequently, we see already that the self--adjoint extension with $\rho\to \infty$ and $(\theta_1,\theta_2)=(0,0)$ contains at least an unstable mode of $H_+$. As in the free case, the ${\cal A}{\cal A}^\dagger$ form of the Hamiltonian does not guarantee, by itself, the positivity of the spectrum in an interval. For $\mu=0$ the two wavefunctions $\psi_\pm^{(0)}$ coincide, while for $\mu \geq 1$ only $\psi_+^{(0)}$ is a normalizable zero mode.

The preceding discussion indicates that, in order to discuss the general solution, one must distinguish a few cases.
\begin{itemize}
    \item  If $0<\mu<1$, the general normalizable solution of the Schr\"odinger equations for $H_\pm$
is a linear combination of two Legendre $P$ functions~\cite{tables},
\beq
\Psi \ = \ \sqrt{\sin\left( \frac{\pi\,z}{z_m}\right)} \left\{ A\ P_{\nu}^{-\,\mu}\left[ \cos\left(\frac{\pi\,z}{z_m}\right)\right] \ + \ B\ P_{\nu}^{\mu}\left[ \cos\left(\frac{\pi\,z}{z_m}\right)\right]\right\} \ . \label{eqpsimu}
\eeq
with
\beq
\nu \ = \ - \  \frac{1}{2} \ + \ \sqrt{m^2 \ + \ \left(\mu \,\pm\,\frac{1}{2}\right)^2}  \label{eqnu}
\eeq
for the two Hamiltonians $H_\pm$, so that
\bea
&\mathrm{for \ } H_+ :& \quad m^2 \ = \ \left(\nu\,+\,\mu\,+\,1\right)\left(\nu\,-\,\mu\right) \ , \nonumber \\
&\mathrm{for \ } H_- :& \quad m^2 \ = \ \left(\nu\,+\,\mu\right)\left(\nu\,-\,\mu\,+\,1\right) \ .
\eea
Note that for $H_+$ there are tachyonic modes in the range  $-\,\frac{1}{2} \leq \nu < \mu$, while for $H_-$ they correspond to the range $-\,\frac{1}{2} \leq \nu < \mathrm{max}\left(-\mu,\mu-1\right)$. We have already seen an example of this type, the wavefunction $\psi_-^{(0)}$ of eq.~\eqref{zeromodepsi}, which has $m^2 \ = \ - \ 2\,\mu$ for $H_+$, and therefore
\beq
\nu \ = \ - \ \frac{1}{2} \ + \ \left| \mu \ - \ \frac{1}{2}\right| \ .
\eeq
There can be additional tachyonic modes for $H_\pm$ with~\footnote{The corresponding wavefunctions are often called conical (or Mehler) functions in the literature~\cite{tables}.}
\beq
\nu\ =\ -\ \frac{1}{2}\ +\ i\,x \ , \label{nuimaginary}
\eeq
with $x$ real, which correspond to
\bea
&\mathrm{for \ } H_+ :& \quad m^2 \ = \ - \ x^2 \ - \ \left(\mu\,+\,\frac{1}{2}\right)^2 \ , \nonumber \\
&\mathrm{for \ } H_- :& \quad m^2 \ = \ - \ x^2 \ - \ \left(\mu\,-\,\frac{1}{2}\right)^2 \ .
\eea

The limiting behavior of $\Psi$ close to the $z=0$ end is~\cite{tables}
\beq
\Psi \ \sim \ \sqrt{2} \left[ \frac{A}{\Gamma(1 + \mu)} \left(\frac{\pi z}{2 z_m}\right)^{\frac{1}{2}+\mu} + \  \frac{B}{\Gamma(1 - \mu)} \left(\frac{\pi z}{2 z_m}\right)^{\frac{1}{2}- \mu}\right] \ , \label{0lim}
\eeq
so that the two coefficients in eq.~\eqref{b.11} are
\beq
C_1 \ = \  \frac{A \,\sqrt{2}}{\Gamma(1+\mu)} \left(\frac{\pi}{2} \right)^{\frac{1}{2}+\mu}\ , \qquad
C_2 \ = \  \frac{B\,\sqrt{2}}{\Gamma(1 - \mu)} \left(\frac{\pi}{2}\right)^{\frac{1}{2}- \mu} \ .
\eeq

The behavior near the other end can be deduced from the connection formulas contained in~\cite{tables},
and in particular from
\bea
P_{\nu}^{\mu}\left[ \cos\left(\frac{\pi\,z}{z_m}\right)\right] \!\!\!&=& \!\!\!  - \ \frac{\sin(\pi\,\nu)}{\sin(\pi\,\mu)}\, P_{\nu}^{\mu}\left[ \cos\left(\pi- \frac{\pi z}{z_m}\right)\right] \nonumber \\ &+& \!\!\!  \frac{\Gamma(\nu+\mu+1) \ \sin \pi(\nu+\mu)}{\Gamma(\nu-\mu+1) \ \sin \pi \mu} \, P_{\nu}^{-\, \mu}\left[ \cos\left(\pi- \frac{\pi z}{z_m}\right)\right]
\eea
and the corresponding one with $\mu$ replaced by $-\mu$. One can thus conclude that
\bea
C_3 \!\!&=&\!\!   \frac{1}{\sin \pi \mu}  \left\{ C_1 \, \sin(\pi \nu) \, + \, C_2\left(\frac{\pi}{2}\right)^{2\mu}  \frac{\Gamma(1-\mu)\,\Gamma(\nu+\mu+{1}) \, \sin\pi(\nu+\mu)}{\Gamma(1+\mu)\,\Gamma(\nu-\mu+{1})}\right\}\ ,  \\
C_4 \!\!&=&\!\!  -\, \frac{1}{\sin \pi \mu} \left\{ C_2 \, \sin(\pi\nu) \, + \, C_1\left(\frac{\pi}{2}\right)^{-\,2\mu} \frac{\Gamma(1+\mu)\,\Gamma(\nu-\mu+{1}) \ \sin \pi(\nu-\mu)}{\Gamma(1-\mu)\,\Gamma(\nu+\mu+{1})}\right\}\ . \nonumber
\eea
Note that the link between $(C_1,C_2)$ and $(C_4,C_3)$ rests on a special SL(2,R) matrix $U\left(\vec{n}{}'\right)$, which is the counterpart of the matrix $V(z_m)$ of eqs.~\eqref{Vmatrix} and \eqref{Vzm} that we obtained in Section~\ref{sec:free} for the free theory. The general self--adjoint boundary condition thus becomes
\beq
U^{-1}\left(\vec{n}\right)\, U\left(\vec{n}{}'\right) \, \underline{C}(0) \ = \ e^{\,-\,i\,\beta}\ \underline{C}(0) \ , \label{eigenvonemu}
\eeq
and the general eigenvalue equation is then determined by
\beq
\mathrm{Tr}\left[U^{-1}\left(\vec{n}\right)\, U\left(\vec{n}{}'\right)\right] \ = \ 2\,\cos\beta \ . \label{sl2rdet}
\eeq
It can be cast in the form
\bea
&& 2\,\cosh\rho \,\sin\theta_1\,\sin\pi\nu \ + \ b(\mu,\nu) \Big(\cosh\rho \,\cos\theta_1 \,+\,\sinh\rho \,\cos\theta_2\Big) \nonumber \\ &-& b(-\mu,\nu)\Big(\cosh\rho \,\cos\theta_1 \,-\,\sinh\rho \,\cos\theta_2\Big) \ = \ 2\,\cos\beta\,\sin\pi\mu \ , \label{3.95}
\eea
where
\beq
b(\mu,\nu) \ = \ \left(\frac{\pi}{2}\right)^{2\mu}  \frac{\Gamma(1-\mu)\,\Gamma(\nu+\mu+{1})}{\Gamma(1+\mu)\,\Gamma(\nu-\mu+{1})} \, \sin\pi\left(\nu+\mu\right) \ .
\eeq
Eq.~\eqref{3.95} determines the mass spectrum for the different self--adjoint extensions. It is manifestly real for real values of $\nu$, but one is also interested in the line~\eqref{nuimaginary}, where it is convenient to recast it in a manifestly real form. To this end note that
\beq
B(\mu,x)\ = \ \frac{b\left(\mu,-\,\frac{1}{2}\,+\,i\,x\right)}{\cosh \pi x} \  = \ - \ \frac{\pi\  \left(\frac{\pi}{2}\right)^{2\mu} \Gamma(1-\mu)}{\Gamma(1+\mu)\,\cosh \pi x\left|\Gamma\left(\frac{1}{2}\,-\,\mu\,+\,i\,x\right)\right|^2} \ ,
\eeq
which is manifestly real. As a result, along the line $\nu = - \frac{1}{2}+i x$ the eigenvalue equation becomes
\bea
&-& 2\,\sin\theta_1 \ + \ B(\mu,x) \Big(\cos\theta_1 \,+\,\tanh\rho \,\cos\theta_2\Big) \nonumber \\ &-& B(-\mu,x)\Big(\cos\theta_1 \,-\,\tanh\rho \,\cos\theta_2\Big) \ = \ \frac{2\,\cos\beta\,\sin\pi\mu}{\cosh \pi x\,\cosh\rho} \ . \label{3.95x}
\eea

In the large--$\rho$ limit the eigenvalue equation simplifies, and can be cast in the form
\beq
2\,\sin\theta_1\,\sin\pi\nu \, + \, b(\mu,\nu)\, \Big(\cos\theta_1 \,+\,\cos\theta_2\Big) \, - \,   b(-\mu,\nu)\,\Big(\cos\theta_1 \,-\,\cos\theta_2\Big) \,=\, 0  \label{3.96}
\eeq
for real values of $\nu$, and as
\beq
\ - \ 2\,\sin\theta_1 \ + \ B(\mu,x) \Big(\cos\theta_1 \,+\,\cos\theta_2\Big) \ - \  B(-\mu,x)\Big(\cos\theta_1 \,-\,\cos\theta_2\Big) \ = \ 0  \label{3.95x2}
\eeq
along the line $\nu = - \frac{1}{2}+i x$.
\begin{figure}[ht]
\centering
\begin{tabular}{cc}
\includegraphics[width=50mm]{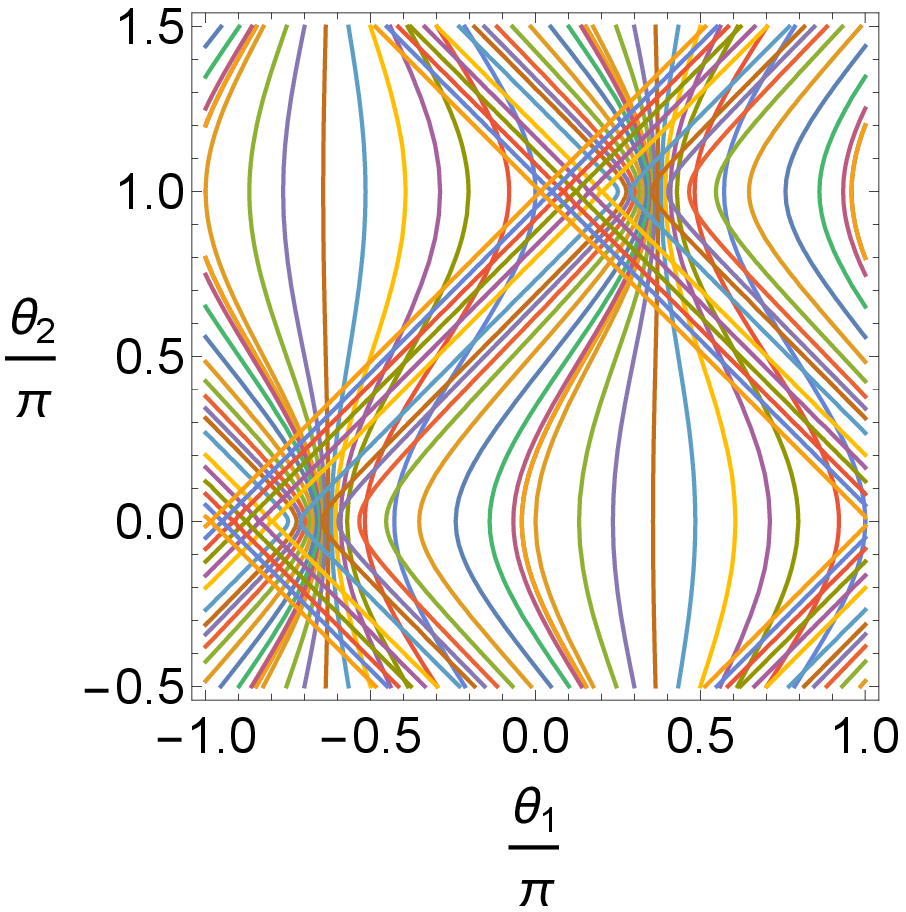}  \qquad\qquad &
\includegraphics[width=50mm]{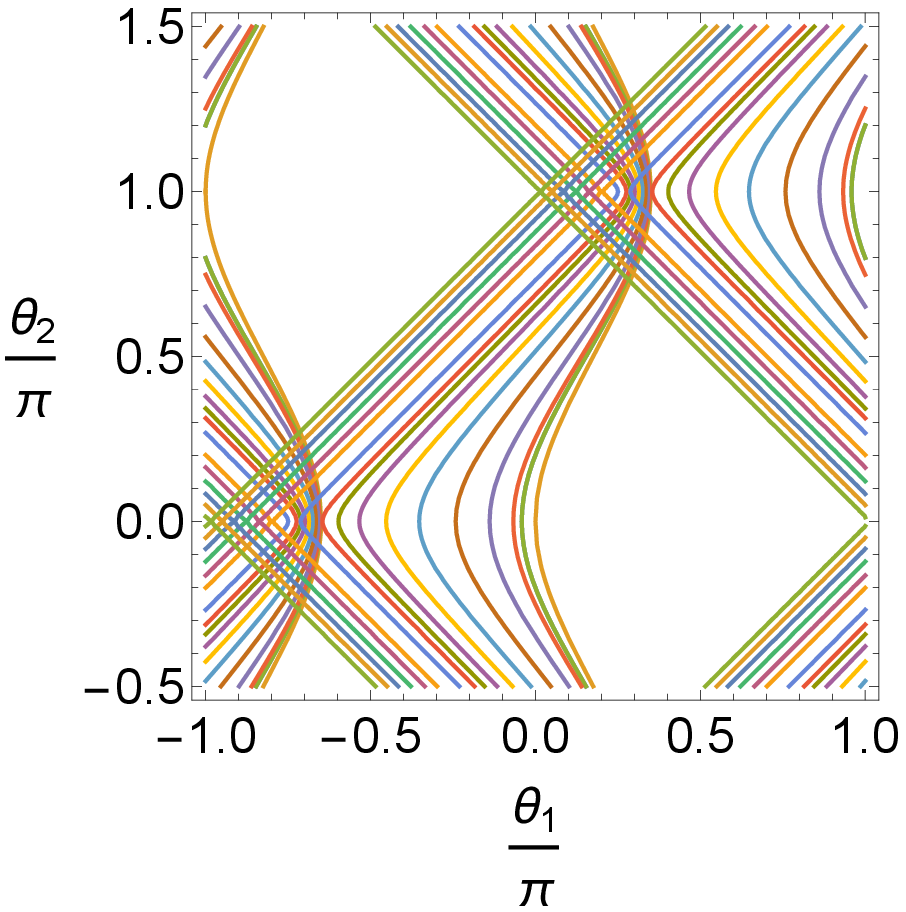} \\
\end{tabular}
\caption{\small For $\mu=\frac{3}{8}$, there are instabilities for $H_+$, for all values of $\theta_1$ and $\theta_2$ away from the special points $(\pi,0)$ and $(0,\pi)$  (left panel), while for $H_-$ there are wide stability regions that include those points (right panel).}
\label{fig:instabilities_nu_real}
\end{figure}

In some special cases, the spectra emerging from eq.~\eqref{3.95} can be exactly determined.
\begin{itemize}
\item If $\theta_1=\pm\,\frac{\pi}{2}$ and $\theta_2=\pm \frac{\pi}{2}$ the eigenvalue equation reduces to
\beq
\sin \pi\nu \ = \ \frac{\cos\beta\,\sin\pi\mu}{\sin\theta_1\,\cosh\rho} \ ,
\eeq
and the solutions correspond to real values of $\nu \geq - \frac{1}{2}$ and read
\beq
\nu \ = \ \nu_0 \ + \ 2\, n \ , \qquad \nu \ = \ - \, \nu_0 \ + \ 2\, n \,+\, 1 \ , \quad n=0,1,\ldots \ ,
\eeq
where
\beq
\nu_0 \ = \ \frac{1}{\pi}\ \arcsin\left(\frac{\cos\beta\,\sin\pi\mu}{\sin\theta_1\,\cosh\rho}\right) \ .
\eeq
There is a tachyon in the first sequence if $\nu_0<0$.
\item For $(\theta_1,\theta_2)=(\pm \pi,0)$ or $(0,\pm \pi)$ the eigenvalue equation reduces to
\beq
b(-\mu,\nu) \ = \ 0 \ ,
\eeq
which is solved by
\beq
\nu \ = \ \mu \ + \ n \ , \qquad  n\,\geq \, 0 \ .
\eeq
The resulting mass spectrum for $H_+$ is
\beq
m^2 \ = \ n\left(2\,\mu\,+1\,+\,n\right) \ , \qquad n=0,1, \ldots \ , \label{m2munu}
\eeq
and there are no tachyons. The spectrum for $H_-$ is obtained adding $2 \mu$ to this result, and
there are also no tachyons along the line~\eqref{nuimaginary}.
This case, which corresponds to $C_2=C_4=0$, is relevant for the ground--state wavefunctions~\eqref{zeromodepsi}.
\item For $(\theta_1,\theta_2)=(0,0)$ or $(\pi,\pi)$ the eigenvalue equation reduces to
\beq
b(\mu,\nu)=0 \ ,
\eeq
which is solved by
\beq
\nu \ = \ - \, \mu \ + \ n \ , \qquad  n\,\geq \, 0 \ .
\eeq
and the resulting mass spectrum for $H_+$
\beq
m^2 \ = \ \left(2\,n\,+1\right)(n - 2 \mu) \ , \qquad n=0,1, \ldots \ , \label{m2munu2}
\eeq
includes one or two tachyons, depending on whether or not $\mu>\frac{1}{2}$, and $H_-$ has one tachyon if $\mu>\frac{3}{4}$.
\item The pattern is similar for generic values of $\theta_1$ and $\theta_2$, and in particular in the large--$\rho$ limit there are typically tachyons in the range $-\frac{1}{2} \leq \nu \leq \mu$ for $H_+$, and fewer for $H_-$ where the upper bound on $\nu$ is reduced to $\mathrm{max}\left(-\mu,\mu-1\right)$. Additional tachyonic modes are present along the line~\eqref{nuimaginary}. For $x$ of order one or larger the term in eq.~\eqref{3.95x2} containing $B(\mu,x)$ becomes dominant, and consequently there is an infinite number of tachyonic modes along the curve
\beq
\cos\theta_1 \ + \ \cos\theta_2 \ = \ 0 \ ,
\eeq
or equivalently along the square $\theta_1 = \pm\,\theta_2 \,\pm\,\pi$,
so that with these special boundary conditions the spectrum is unbounded from below. Different curves
of constant tachyon mass for the case $\mu=\frac{3}{8}$ are displayed in fig.~\ref{fig:instabilities_nu_real}.
\end{itemize}

\item For $\mu=0$ the two functions used in eq.~\eqref{eqpsimu} are no longer independent, and consequently one must start from~\cite{tables}
\beq
\Psi \ = \ \sqrt{\sin\left( \frac{\pi\,z}{z_m}\right)} \left\{ A\ P_{\nu}^{0}\left[ \cos\left(\frac{\pi\,z}{z_m}\right)\right] \ + \ B\ Q_{\nu}^{0}\left[ \cos\left(\frac{\pi\,z}{z_m}\right)\right]\right\} \ , \label{psimu0}
\eeq
with associated Legendre functions of the first and second kind.

The limiting behavior close to $z=0$ is then~\cite{tables}
\beq
\Psi \ = \ \left( \frac{\pi\,z}{z_m}\right)^\frac{1}{2} \ A \ - \ B \left( \frac{\pi\,z}{z_m}\right)^\frac{1}{2} \left[\log\left(\frac{z}{z_m}\right) \ + \ \sigma(\nu) \right] \ ,
\eeq
where
\beq
\sigma(\nu) \ = \ \log\left(\frac{\pi}{2}\right) \ - \ \psi(1) \ + \ \psi\left(\nu + 1\right)  \ , \label{sigma}
\eeq
and $\psi(\nu) = \frac{\Gamma'(\nu)}{\Gamma(\nu)}$,  with $\psi(1) = -\, \gamma \simeq - \,0.77$, the Euler--Mascheroni constant, while the limiting behavior close to the other end is determined by the connection formulae~\cite{tables}
\bea
P_{\nu}^{0}\!\!\left[ \cos\left(\frac{\pi\,z}{z_m}\right)\right]\!\!\!\!\! &=&\!\!\! \cos\pi \nu \, P_{\nu}^{0}\!\!\left[ \cos\left(\pi  -  \frac{\pi\,z}{z_m}\right)\right] - \frac{2\,\sin \pi \nu}{\pi}\,  Q_{\nu}^{0}\!\!\left[ \cos\left(\pi - \frac{\pi\,z}{z_m}\right)\right]  \\
Q_{\nu}^{0}\!\!\left[ \cos\left(\frac{\pi\,z}{z_m}\right)\right]\!\!\!\!\! &=&\!\!\! -\,\cos\pi \nu \, Q_{\nu}^{0}\!\!\left[ \cos\left(\pi - \frac{\pi\,z}{z_m}\right)\right] - \frac{\pi\,\sin \pi \nu}{2}\,  P_{\nu}^{0}\left[ \cos\left(\pi - \frac{\pi\,z}{z_m}\right)\right] , \nonumber
\eea
which give
\bea
\Psi &=& \sqrt{\sin\left( \frac{\pi\,z}{z_m}\right)} \, P_{\nu}^{0}\left[ \cos\left(\pi \ - \ \frac{\pi\,z}{z_m}\right)\right] \left( A \,\cos \pi \nu \ - \ B \ \frac{\pi\,\sin \pi \nu}{2} \right) \nonumber \\
&+&
\sqrt{\sin\left( \frac{\pi\,z}{z_m}\right)} \, Q_{\nu}^{0}\left[ \cos\left(\pi \ - \ \frac{\pi\,z}{z_m}\right)\right] \left( - A \ \frac{2\,\sin \pi \nu}{\pi} \ - \  B \,\cos \pi \nu  \right)  \ .
\eea

Consequently
\beq
C_1 \ = \  A\ {\pi}^\frac{1}{2} \ - \  B \ {\pi}^\frac{1}{2} \ \sigma(\nu) \ , \qquad
C_2 \ = \  - \ B \ {\pi}^\frac{1}{2} \ ,
\eeq
while
\bea
C_4 \!\!\!&=&\!\!\! \frac{2\,\sin \pi \nu}{\pi}\,C_1 \ - \ \left[\cos\pi\nu \ + \ \frac{2}{\pi}\,\sigma(\nu)\,\sin\pi\nu \right]  C_2 \ , \nonumber \\
C_3 \!\!\!&=&\!\!\! C_1 \left[ \cos \pi \nu \,+\, \frac{2}{\pi}\,\sigma(\nu)\,\sin \pi \nu  \right] \nonumber \\ &+& C_2 \left[\frac{\pi}{2}\,\sin \pi \nu \, -\, \frac{2}{\pi}\,\sigma(\nu)^2\,\sin \pi \nu  \,-\,2\,\sigma(\nu)\cos\pi\nu \right]\ . \label{eqsU}
\eea
Note that these transformations define, once more, an $SL(2,R)$ matrix $U\left(\vec{n}{}'\right)$. The general eigenvalue equation is then obtained from eq.~\eqref{sl2rdet}, and reads
\bea
&&\left(\cosh\rho\,\cos\theta_1\,+\,\sinh\rho\,\cos\theta_2\right)\left[\sin\pi\nu \left(\frac{\pi}{4} \,-\,\frac{1}{\pi}\, \sigma^2(\nu)\right) \ - \ \sigma(\nu)\,\cos\pi\nu \right] \nonumber \\
&&+\left(\cosh\rho\,\cos\theta_1\,-\,\sinh\rho\,\cos\theta_2\right) \frac{\sin \pi\nu}{\pi} \nonumber \\
&&+ \ \cosh\rho\,\sin\theta_1\left[ \frac{2}{\pi}\, \sin\pi\nu\, \sigma(\nu) \ + \ \cos\pi\nu \right] \ = \ \cos\beta \ ,
\eea
and in this case eq.~\eqref{eqnu} gives
\beq
m^2 \ = \ \nu(\nu+1) \ . \label{eingevmu0}
\eeq
Massless modes obtain, in this sector with $\mu=0$, if there are solutions with $\nu=0$.
\begin{figure}[ht]
\centering
\includegraphics[width=60mm]{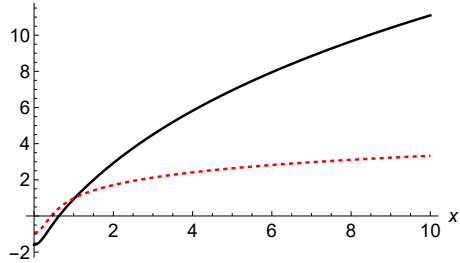}
\caption{\small The function $f_1(x)$  (black, solid) and the function $f_2(x)$  (red, dashed).}
\label{fig:positivity_sin_lower}
\end{figure}

As for $0<\mu<1$, tachyonic modes lie in the real interval $- \frac{1}{2} \leq \nu <0$, which corresponds to solutions with $m^2 \geq - \, \frac{1}{4}$, and also in another interval, with $\nu\, =\, - \,\frac{1}{2} \, + \, i \,x$, which corresponds to solutions with $m^2 = \,-\, \frac{1}{4} \,-\, x^2$, with $x$ real. A special property of the $\psi$ function~\cite{tables},
\beq
Im\left[\psi\left(\frac{1}{2}\,+\, i\,x\right)\right] \ = \ \frac{\pi}{2}\,\tanh\left(\pi x\right)  \ ,
\eeq
makes the eigenvalue equation~\eqref{eingevmu0} real also along the additional line, where it becomes
\bea
&&\left(\cosh\rho\,\cos\theta_1\,+\,\sinh\rho\,\cos\theta_2\right)\,f_1(x) \, - \,\left(\cosh\rho\,\cos\theta_1\,-\,\sinh\rho\,\cos\theta_2\right) \nonumber \\ &&-\   2\, \cosh\rho\,\sin\theta_1 \,f_2(x) \ = \ - \  \pi\ \frac{\cos\beta}{\cosh\pi x} \ , \label{tachyon_eqx}
\eea
where
\beq
f_1(x) \ = \ \Re\left[\sigma\left(-\,\frac{1}{2}+ i x\right)\right]^2\,-\,\frac{\pi^2}{4\,\cosh^2 \pi\,x}  \ , \quad f_2(x) \ = \ \Re\left[\sigma\left(-\,\frac{1}{2}+ i x\right)\right] .
\eeq
There are generally tachyons in this sector, whose squared masses lie below $-\,\frac{1}{4}$.
These two functions are displayed in fig.~\ref{fig:positivity_sin_lower}: notice how $f_1(x)$ soon becomes much larger than $f_2(x)$, so that the spectrum is actually unbounded from below on the surface
\beq
\cosh\rho\,\cos\theta_1\,+\,\sinh\rho\,\cos\theta_2  \ = \ 0 \ .
\eeq

\begin{figure}[ht]
\centering
\begin{tabular}{cc}
\includegraphics[width=55mm]{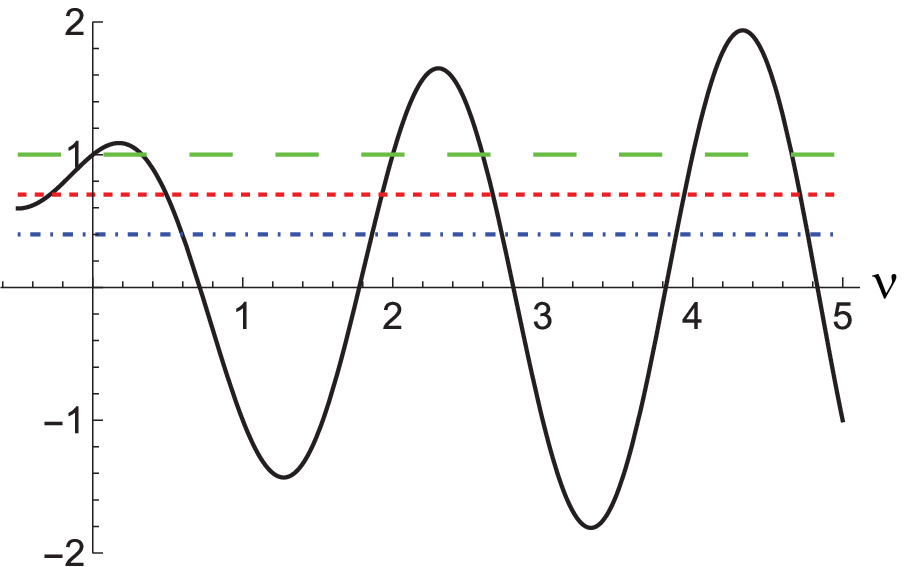} \qquad\qquad &
\includegraphics[width=55mm]{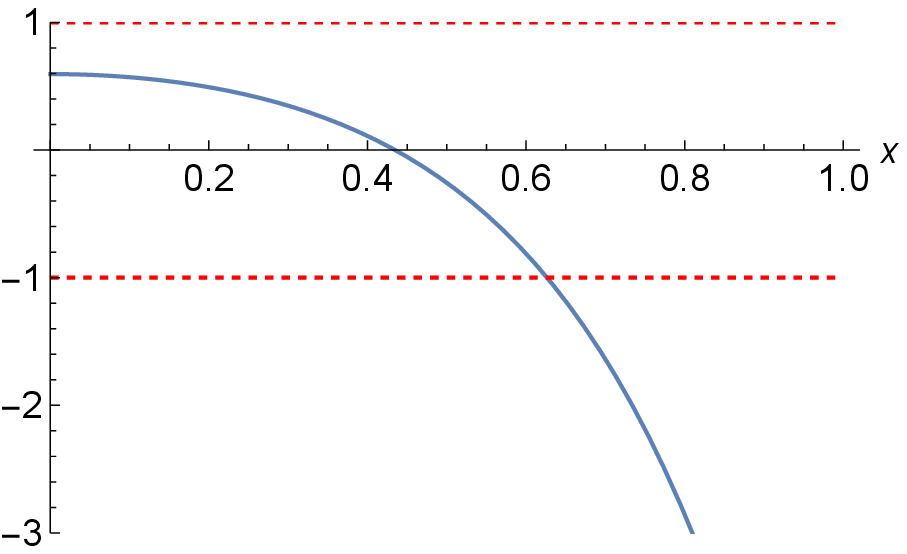} \\
\end{tabular}
\caption{\small Left panel: the simplified eigenvalue equation \eqref{eigenvb} for $\nu$ real and $\nu > - \,\frac{1}{2}$, with $\cos\beta=1$ (solid vs green,long-dashed), $\cos\beta=0.7$ (solid vs red,dashed) and $\cos\beta=0.4$ (solid vs blue, dash-dotted). Right panel: the same eigenvalue equation for $\nu = - \frac{1}{2} \ + \ i x$.}
\label{fig:mu01}
\end{figure}

The eigenvalue equation simplifies in some special cases:
\begin{itemize}
    \item If $\rho=0$ and $\theta_1=\frac{\pi}{2}$, it reduces to
    \beq
\frac{2}{\pi}\, \sin\pi\nu\, \sigma(\nu) \ + \ \cos\pi\nu \ = \ \cos\beta \ , \label{eigenvb}
    \eeq
    which can be solved graphically for real values of $\nu$, as in the left panel of fig.~\ref{fig:mu01}. Note the presence of a zero mode for $\beta=0$ and the emergence of a tachyonic mode for $0 < \beta < \frac{\pi}{3}$. These boundary conditions actually correspond to $U=-\,i\,\sigma_2$, and eqs.~\eqref{eqsU} show that if $\beta=0$ the spectrum is given exactly by $\nu=2 n$, with $n=0,1,\ldots$, as can be foreseen from the left panel of fig.~\ref{fig:mu01}. The right panel then shows the absence of tachyons along the line~\eqref{nuimaginary}, so that the system is completely stable with these boundary conditions, which give
    \beq
    C_4 \ = \ - \ C_2 \ , \qquad C_3 \ = \ C_1  \ - \ 2\, \sigma(2n)\, C_2 \ ,
    \eeq
    while compatibility with the $U$ matrix demands that $C_2=C_4=0$.
    There is a limited range of values of $x$ for which the eigenvalue equation can be solved by $\nu = - \frac{1}{2}+i x$, as shown in the right panel of fig.~\ref{fig:mu01}, with an upper bound for $x$ of order one. These modes have tachyonic masses with $m^2 = -\frac{1}{4} - x^2 > - \frac{5}{4}$. Actually, no tachyons are present altogether if $\beta < \frac{\pi}{3}$, as can be seen from fig.~\ref{fig:mu01}.
\item In the $\rho \to \infty$ limit eq.~\eqref{eingevmu0} reduces to
\bea
&&\left(\cos\theta_1\,+\,\cos\theta_2\right)\left[\sin\pi\nu \left(\frac{\pi}{4} \,-\,\frac{1}{\pi}\, \sigma^2(\nu)\right) \ - \ \sigma(\nu)\,\cos\pi\nu \right] \label{mu0nuerhoinf} \\
&&+\left(\cos\theta_1\,-\,\cos\theta_2\right) \frac{\sin \pi\nu}{\pi} \ + \ \sin\theta_1\left[ \frac{2}{\pi}\, \sin\pi\nu\, \sigma(\nu) \ + \ \cos\pi\nu \right] \ = \ 0 \ , \nonumber
\eea
to be considered for $\nu \geq - \frac{1}{2}$, while eq.~\eqref{tachyon_eqx} reduces to
\beq
\left(\cos\theta_1\,+\,\cos\theta_2\right)\,f_1(x) \, - \,\left(\cos\theta_1\,-\,\cos\theta_2\right) \ - \   2\,\sin\theta_1 \,f_2(x) \ = \ 0 \ . \label{tachyon_eqx_rhoinf}
\eeq
\item If $\theta_1=0$, along the $x$ line the tachyonic eigenvalue equation~\eqref{tachyon_eqx_rhoinf}  reduces to
\beq
f_1(x) \ = \ \tan^2\left(\frac{\theta_2}{2}\right) \ ,
\eeq
and as one can see from the left panel of fig.~\ref{fig:positivity_sin_lower} it always has  solutions as $\theta_2$ spans the $[0,\pi]$ range, with values of $x$ ranging from about 0.6 to $\infty$. These tachyonic modes can thus have arbitrarily negative values of $m^2$. In a similar fashion, if $\theta_1=\pi$ the eigenvalue equation~\eqref{tachyon_eqx_rhoinf} reduces to
\beq
f_1(x) \ = \ \cot^2\left(\frac{\theta_2}{2}\right) \ ,
\eeq
and as $\theta_2$ spans the $[0,\pi]$ range, $x$ varies from $\infty$ to about 0.6.
If $\theta_1 \neq 0,\pi$ the equation takes the form
\beq
\frac{\cos\theta_1\,+\,\cos\theta_2}{2\, \sin\theta_1}\,f_1(x) \, - \,\frac{\cos\theta_1\,-\,\cos\theta_2}{2\, \sin\theta_1} \ - \  \,f_2(x) \ = \ 0 \ , \label{tachyon_eq2}
\eeq
which can have zero, one or two solutions, as one can see graphically. The first type of solutions, which correspond to stable vacua, obtain when the first coefficient is positive and not too small and the second is negative and not too small.

\item There are special choices of boundary conditions for which eq.~\eqref{mu0nuerhoinf} is exactly solvable, $\left(\theta_1,\theta_2\right)=\left(0,\pi\right)$ or $\left(\pi,0\right)$. They correspond to
\beq
C_2 \ = \ C_4\ = \ 0 \ ,  \label{c24}
\eeq
and in both cases the eigenvalue equation reduces to
\beq
\sin \pi \nu \ = \ 0 \ ,
\eeq
so that
\beq
m^2 \ = \ n\left(n+1\right) \ , \qquad n=0,1,\ldots \ . \label{spectrum_graviton}
\eeq
There are no tachyons even along the additional line, since eq.~\eqref{tachyon_eqx_rhoinf} admits no solutions with these choices of parameters.
\begin{figure}[ht]
\centering
\begin{tabular}{cc}
\includegraphics[width=60mm]{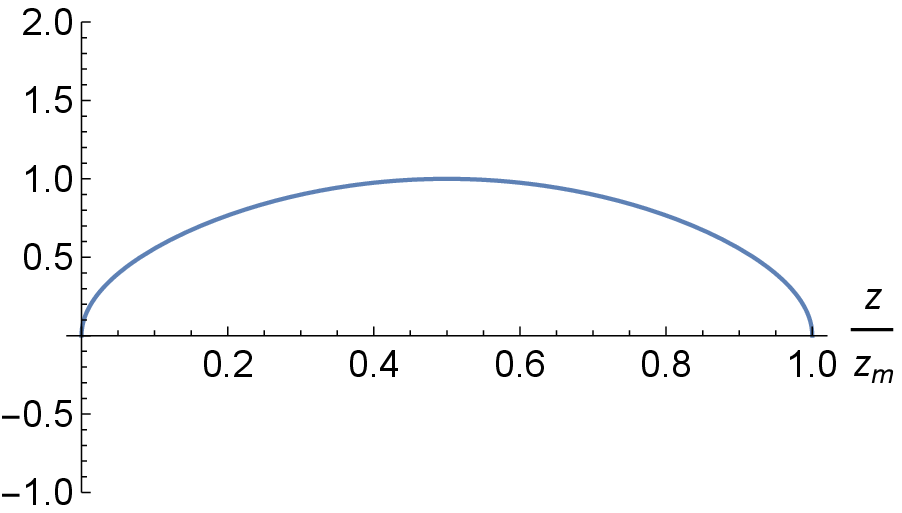} \qquad \qquad &
\includegraphics[width=60mm]{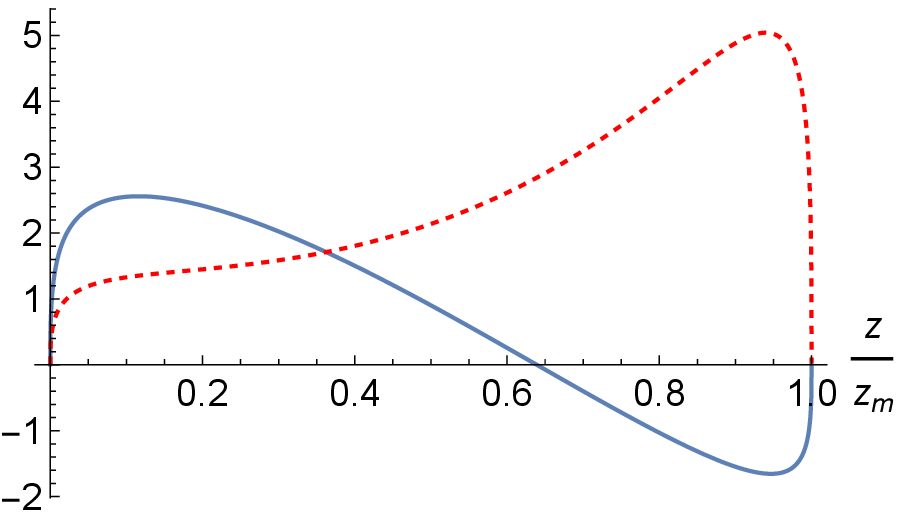} \\
\end{tabular}
\caption{\small The (non--normalized) zero--mode wavefunctions of eq.~\eqref{zeromode12} (blue, solid, left panel) and of eq~\eqref{zeromodelog} (blue, solid, right panel), and the non--normalized wavefunction of the corresponding tachyonic mode (red, dashed, right panel).}
\label{fig:two_wavefunctions}
\end{figure}
\end{itemize}

With the preceding choices of self--adjoint boundary conditions leading to eq.~\eqref{c24}, the zero mode is the simple solution in eq.~\eqref{zeromodepsi} for $\beta=\frac{1}{2}$, proportional to
\beq
\psi \ = \ \sqrt{\sin\left(\frac{\pi\,z}{z_m}\right) } \ . \label{zeromode12}
\eeq

The other independent solution in eq.~\eqref{psimu0}, which corresponds to $C_1=0$, is proportional to
\beq
\psi(z) \ = \ \sqrt{\sin\left(\frac{\pi z}{z_m}\right)} \ \log\left[ \frac{\pi}{2}\, \cot\left(\frac{\pi z}{2\,z_m}\right)\right]^2 \ , \label{zeromodelog}
\eeq
and has $C_3= -\,2\,\log\left(\frac{\pi}{2}\right) C_2$, $C_4 = \,-\,C_2$. Using eq.~\eqref{cond_sing1n}, one can see that these choices correspond to
\beq
\theta_1 \ = \ \theta_2 \ = \ \arctan\left[ 2 \log\left( \frac{\pi}{2} \right) \right] \ . \label{theta12bc}
\eeq
The spectrum resulting from these boundary conditions can be deduced from fig.~\ref{fig:eigenvalues12}, where it corresponds to the intersections of the dashed curves with the real axis. In particular, the left panel refers to the real range for $\nu$, while the right panel identifies a single tachyonic eigenvalue along the complex line~\eqref{nuimaginary}, with $x \simeq 1$.
The wavefunction~\eqref{zeromodelog} has a node, as can be seen in fig.~\ref{fig:two_wavefunctions}, and indeed there is one lower eigenvalue with these self--adjoint boundary conditions, as we have just seen. The wavefunction of the tachyonic mode is a Mehler function, and is displayed in fig~\ref{fig:two_wavefunctions}: it has no nodes in the interior of the interval, and is orthogonal to the zero mode, as expected.
\begin{figure}[ht]
\centering
\begin{tabular}{cc}
\includegraphics[width=55mm]{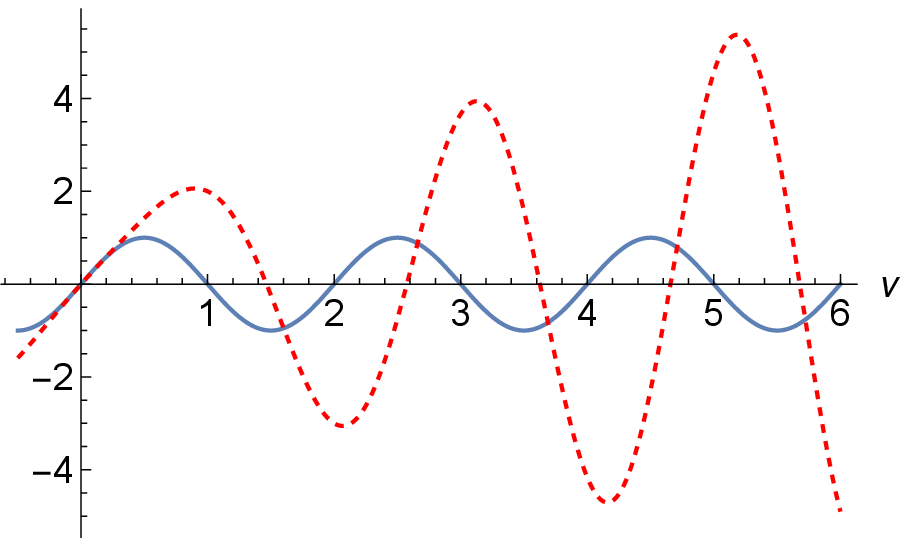} \qquad\qquad &
\includegraphics[width=55mm]{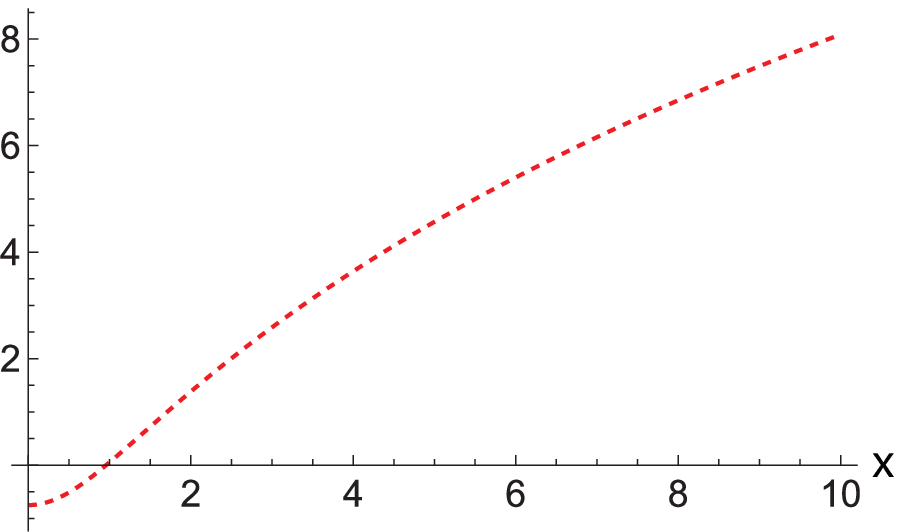} \\
\end{tabular}
\caption{\small In the left panel, the intersections with the horizontal axis identify the spectrum of $\nu$ corresponding to the boundary conditions of the zero--mode wavefunctions in eqs.~\eqref{zeromode12} (solid), and in eq.~\eqref{zeromodelog} (dashed). In the right panel, the intersection with the horizontal axis identifies the value of $x$ corresponding to the tachyonic mode with the boundary conditions~\eqref{theta12bc}.}
\label{fig:eigenvalues12}
\end{figure}

In fact, the presence of tachyons is generic for $\mu=0$, as can seen from fig.~\ref{fig:instabilities_mu0}, which collects the lines of constant tachyonic mass in this case. Consequently, we are led to conclude that the choice $C_2=C_4=0$, which results in the simple spectrum~\ref{spectrum_graviton}, is \emph{the only self--adjoint boundary condition that is free of instabilities for $\mu=0$}, in the $\rho \to \infty$ limit that, as we have seen, translates into independent boundary conditions at the ends of the interval.
\begin{figure}[ht]
\centering
\includegraphics[width=60mm]{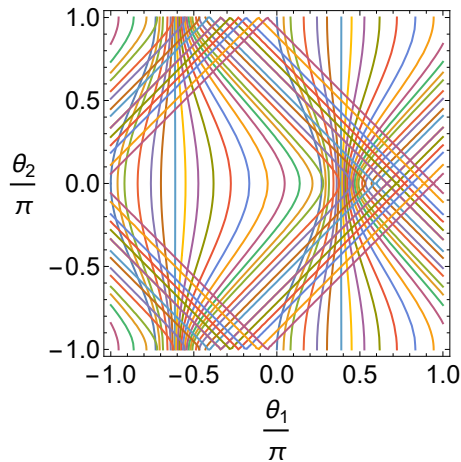}
\caption{\small For $\mu=0$, there are instabilities for all values of $\theta_1$ and $\theta_2$ away from the special point $(\pi,0)$, which corresponds to the spectrum of eq.~\eqref{spectrum_graviton} that starts with a massless ``graviton'' mode with the wavefunction~\eqref{zeromode12}.}
\label{fig:instabilities_mu0}
\end{figure}
\item For $\mu=1$, the value of interest for scalar perturbations in the orientifold 9D vacuum, only one choice of wavefunction,
\beq
\psi \ = \ A \, \sqrt{\sin\left( \frac{\pi\,z}{z_m}\right)}\, P_{\nu}^{-\,1}\left[ \cos\left(\frac{\pi\,z}{z_m}\right)\right] \ ,
\eeq
is compatible with the $L^2$ condition at the left end of the interval. The behavior at the right end is then determined by the connection formula~\cite{tables}
\bea
P_{\nu}^{-\,1}\left[ \cos\left( \frac{\pi\,z}{z_m}\right)\right] &=& -\,\cos\pi\nu \, P_{\nu}^{-\,1}\left[ \cos\pi\left(1 -\frac{z}{z_m}\right)\right]\nonumber \\ &+& \frac{2}{\pi}\, \sin\pi\nu\ Q_{\nu}^{-\,1}\left[ \cos\pi\left(1 - \frac{z}{z_m}\right)\right]
\eea
in the limit $z \to z_m$. Consequently, if the coefficient of $Q_{\nu}^{-\,1}$ does not vanish,
\beq
\psi \ \sim \ \frac{1}{\sqrt{\left(1 - \frac{z}{z_m}\right)}} \ ,
\eeq
as $z\to z_m$, since
\beq
Q_{\nu}^{-\,1} \left[ \cos\pi\left(1 - \frac{z}{z_m}\right)\right]  \ \sim \ \frac{1}{\pi\,\nu(\nu+1) \left(1 - \frac{z}{z_m}\right)} \ , \label{3.111}
\eeq
and the $L^2$ condition thus demands that
\beq
\frac{\sin \pi\nu}{\pi \nu(\nu+1)} \ = \ 0 \ .
\eeq
\begin{figure}[ht]
\centering
\includegraphics[width=60mm]{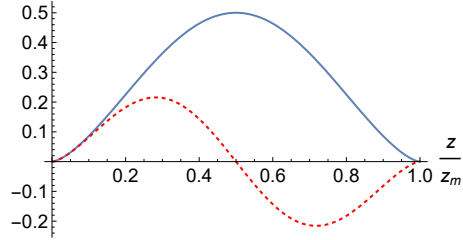}
\caption{\small The non--normalized ground--state wavefunction for $\mu=1$ (blue, solid) and the corresponding one for the first excited state (red, dashed).}
\label{fig:two_wavefunctionsmu1}
\end{figure}

Combining this condition with eq.~\eqref{eqnu}, one can conclude that
\beq
\nu \ \in \mathbb{N} \ = \ 1, 2, 3, \ldots \ .
\eeq
The corresponding $m^2$ eigenvalues are then
\bea
m^2 &= \ (n-1)(n+2) \ = \ 0, 4, 10, \ldots \qquad & \mathrm{for \ H_+} \ , \nonumber \\
m^2 &= \quad n(n+1) \qquad \ = \ 2, 6, 12, \ldots \qquad & \mathrm{for \ H_-} \ . \label{scalar_spectra}
\eea

There is a normalizable zero mode, given in eq.~\eqref{zeromode}, only for $H_+$, which is displayed in fig.~\ref{fig:two_wavefunctionsmu1} together the wavefunction of the first excited state, and there are no tachyons in both cases. These spectra differ by a constant shift in $m^2$, as is manifest in eq.~\eqref{pot_sin}, and the spectrum for $H_-$ is identical to the massive one obtained for $\mu=0$ in eq.~\eqref{spectrum_graviton}, which corresponds to the special boundary conditions for which the ground--state wavefunction is annihilated by ${\cal A}^\dagger$.
\end{itemize}

\subsubsection{\sc The Issue of Positivity} \label{sec:positivity}

In Section~\ref{sec:sing_pot} we have recast the spectral problems for the different sectors of the 9D compactifications of~\cite{dm_vacuum} in terms of Schr\"odinger--like operators of the form
\beq
{\cal A} \, {\cal A}^\dagger \ \psi \ = \ m^2 \ \psi \ , \label{b.282}
\eeq
with
\beq
{\cal A} \ = \ \partial_z \ + \ \omega(z) \ , \qquad {\cal A}^\dagger \ = \ - \  \partial_z \ + \ \omega(z) \ .
\eeq
Similar steps were made in~\cite{bms}.

Multiplying eq~\eqref{b.282} by $\psi^\star$ and integrating gives
\beq
m^2 \int_a^b dz \ \left| \psi\right|^2 \ = \ \int_a^b dz \  \left| {\cal A}^\dagger\ \psi\right|^2 \ + \ \left[ \psi^\star \, {\cal A}^\dagger\,\psi \right]_a^b \ , \label{m2_boundary}
\eeq
and one can thus conclude that if
\beq
\left[ \psi^\star \, {\cal A}^\dagger\,\psi \right]_a^b \ \geq \ 0 \ ,  \label{bc_pos}
\eeq
as $b$ and $a$ approach the singular ends $z_m$ and $0$, then
$m^2 \,\geq\, 0$. If this condition does not hold, positivity is not guaranteed, and as we saw in the preceding section self--adjoint boundary conditions can lead to the emergence of tachyonic modes even if the factorization of the Schr\"odinger operator holds, and in particular in the absence of a potential. The free case was already analyzed in Section~\ref{sec:free_positive}, and here we would like to add some considerations for the singular potentials.

For the cases of interest, as we have seen, close to the left end of the interval
\beq
{\cal A}_\pm \ \sim \ \partial_z \ + \ \frac{\beta_\pm}{z} \ ,
\eeq
and similarly close to the right end
\beq
{\cal A}_\pm \ \sim \ \partial_z \ - \ \frac{\beta_\pm}{z_m \ - \ z} \ .
\eeq

\begin{figure}[ht]
\centering
\begin{tabular}{cc}
\includegraphics[width=50mm]{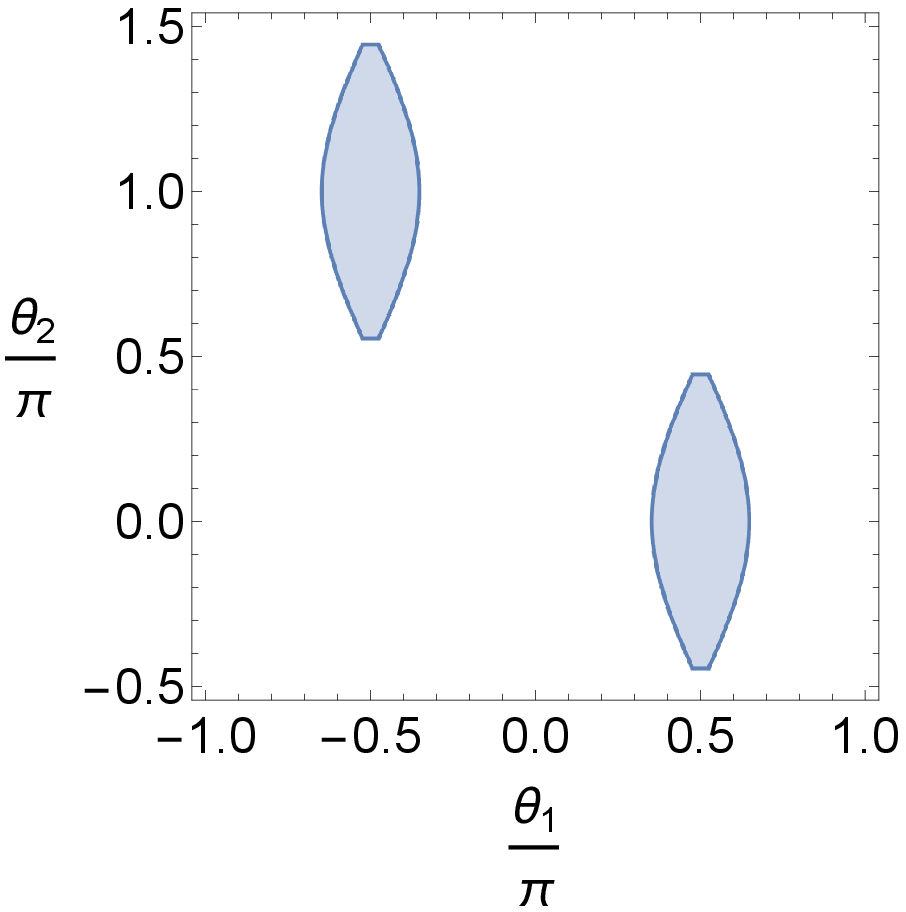} \qquad \qquad &
\includegraphics[width=50mm]{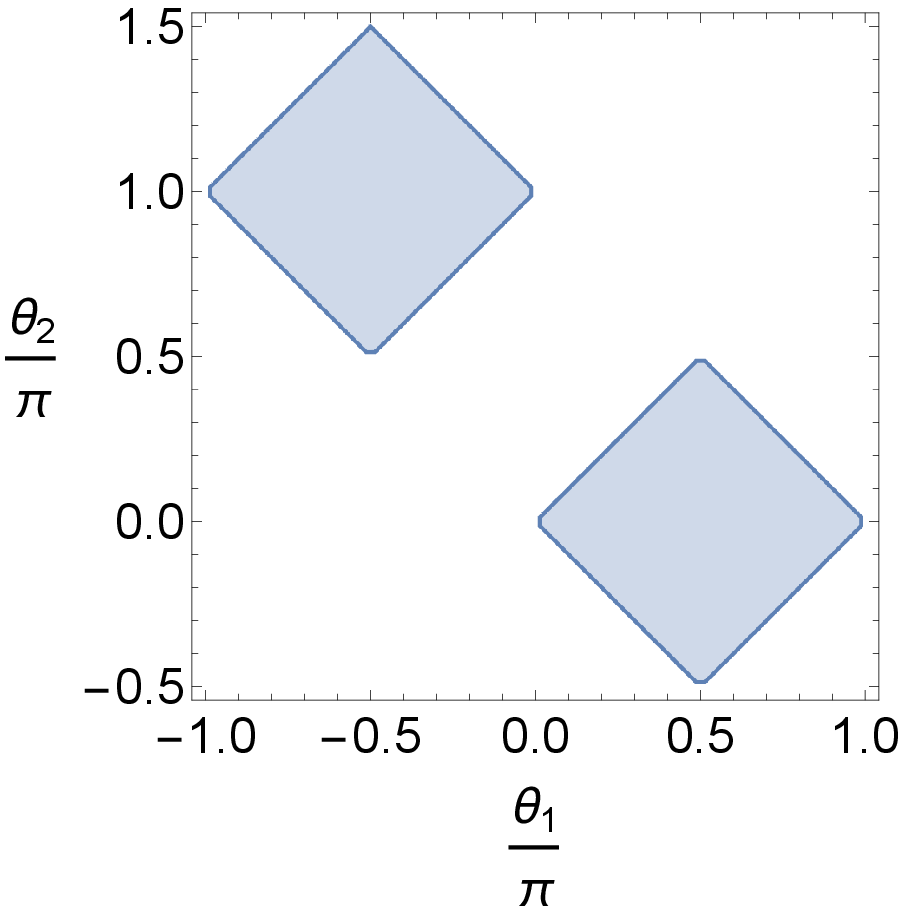} \\
\end{tabular}
\caption{\small The shaded areas are the regions of moduli space where positivity holds for $H_-$, for $\rho=0.2$ (left panel), and for large $\rho$ (right panel).}
\label{fig:positivity_betaminus}
\end{figure}

For $0<\mu<1$, $\left|{\cal A}_+^\dagger\,\psi\right|^2$ is not normalizable, and in that case eq.~\eqref{m2_boundary}
contains pairs of singular contributions as $a$ and $b$ approach the endpoints, while $\left|{\cal A}_-^\dagger\,\psi\right|^2$
is normalizable, and the boundary term for $H_-$ reads
\beq
2\mu \left(C_2^\star\,C_1 \ + \ C_4^\star\,C_3\right) \ .
\eeq
In this case eq.~\eqref{bc_pos} translates into the inequality
\beq
\underline{C}(0)^\dagger \left[\sigma_1 \ + \ U^\dagger\,\sigma_1\, U\right]\underline{C}(0) \geq \, 0 \ ,
\eeq
and using eq.~\eqref{global_ads3_1} this can be turned into the conditions
\bea
&& \sin\left(\theta_1\ + \ \theta_2\right) \ \geq \ 0 \ , \nonumber \\
&&  \cos^2 \theta_1 \ - \ \tanh^2 \rho \, \cos^2 \theta_2  \ \leq \ 0 \ ,  \label{pos_sl2r2}
\eea
which hold in the shaded regions of fig.~\ref{fig:positivity_betaminus}. One can thus conclude that positivity holds for $H_-$ within these regions. This pattern can be recognized in the last two panels of fig.~\ref{fig:instabilities_nu_real}, which refer to $H_-$: the tachyonic curves displayed there lie outside these regions.

Furthermore, eq.~\eqref{delta_H} indicates that, within the same regions of fig.~\ref{fig:positivity_betaminus}, the eigenvalues of $H_+$ are bounded from below, and
\beq
m^2 \geq - 2\,\mu \ .
\eeq
However, the preceding argument does not identify the stability regions of $H_+$.

For $\mu=0$ the situation is similar to what we described for $H_+$, since
\beq
{\cal A}^\dagger \,\psi \ \sim \ \frac{C_2}{\sqrt{z}} \ ,
\eeq
which is not normalizable, and the boundary term is divergent, although the divergent contributions cancel again between boundary and bulk. As a result, the analysis is not conclusive in this case unless $C_2=C_4=0$. Indeed, as we saw in Section~\ref{sec:exact}, there are tachyonic modes for all values of $\theta_1$ and $\theta_2$ away from the special point $(\theta_1,\theta_2)=(\pi,0)$.

Even when the Hamiltonian factorizes as in eq.~\eqref{b.282}, special choices of boundary conditions are thus needed to grant positivity.
In general, however, if there is a normalizable zero mode $\psi_0$ such that
\beq
{\cal A}^\dagger\,\psi_0 \ = \ 0  \ , \label{b.33}
\eeq
one can conclude that positivity holds for the whole spectrum subject to its self--adjoint boundary conditions. In this case,
the ground state is massless and clearly satisfies the preceding condition~\eqref{bc_pos}, while the mass term in eq.~\eqref{b.282} is subdominant close to the singularities present  in our problem at the ends of the interval. The corresponding massive modes are thus of the form
\beq
\psi \ = \ \psi_0 \ + \ \delta\,\psi \ ,
\eeq
and near the origin $\delta\,\psi \sim m^2\, z^{\alpha+2}$ if $\psi_0 \sim z^\alpha$. As a result, the contribution to eq.~\eqref{bc_pos} has the leading behavior $z^{2\alpha+1}$, and vanishes if $\alpha> -\frac{1}{2}$, a condition that must hold if $\psi_0$ is in $L^2$, which grants a positive spectrum. Similar considerations apply close to $z_m$, and consequently in all sectors of this type positivity is guaranteed by the limiting behavior of the zero mode.

\section{\sc Application to Tensor, Scalar and Vector  Perturbations} \label{sec:applications}
\begin{figure}[ht]
\centering
\begin{tabular}{ccc}
\includegraphics[width=48mm]{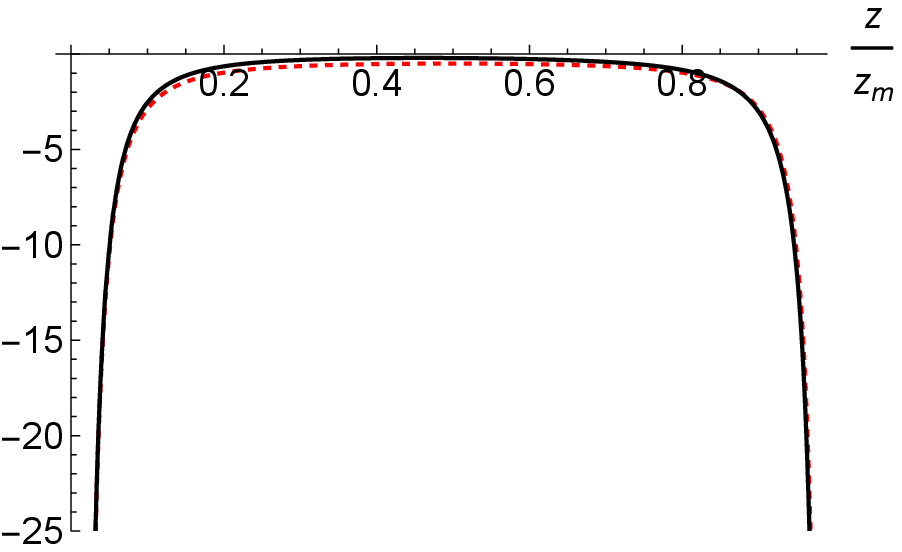} \quad  &
\includegraphics[width=48mm]{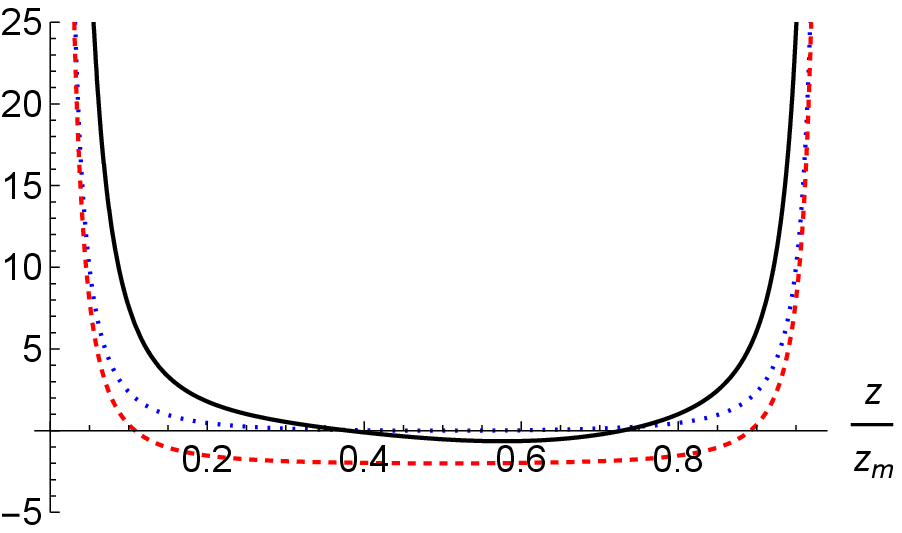} \quad  &
\includegraphics[width=48mm]{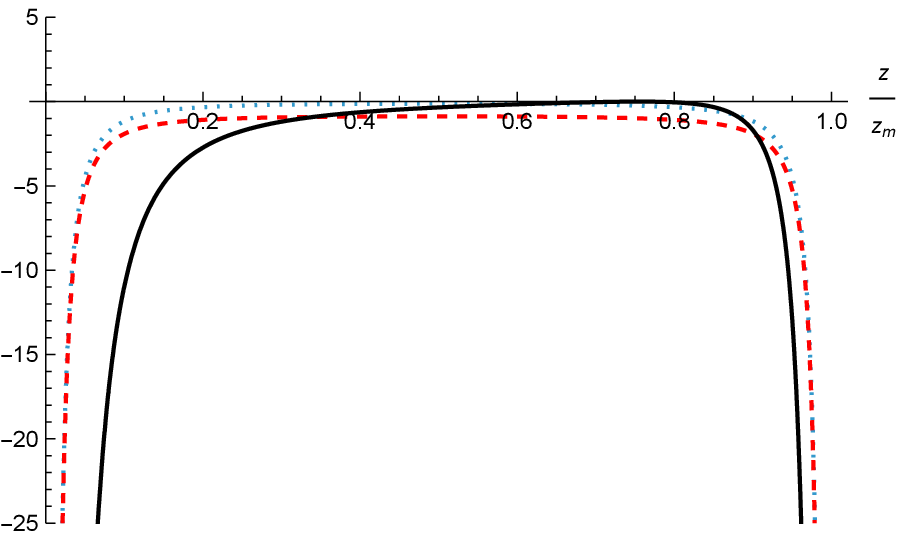} \\
\end{tabular}
\caption{\small The potentials for tensor perturbations (left panel, solid), scalar perturbations (middle panel, solid) and vector perturbations (right panel, solid) for $\gamma=\gamma_c$, in units of $\frac{1}{z_m^2}$ and as functions of $\frac{z}{z_m}$. The potential of eq.~\eqref{pot_sin} with $\mu=0$ is the red dashed curve in the left panel, while the $H_+$ potentials for $\mu=1$ and $\mu=\frac{3}{8}$ are the red dashed curves in the middle and right panels. Finally, the $H_-$ potentials for $\mu=1$ and $\mu=\frac{3}{8}$ are the green dotted curved in the middle and right panels.}
\label{fig:exact_vs_sin}
\end{figure}

We have seen in Section~\ref{sec:sing_pot} that, within the range $\gamma\leq \gamma_c$ that includes the tadpole potentials of the orientifold models, the two parameters $\mu$ and $\tilde{\mu}$ characterizing the singular behavior at the ends of the interval coincide for tensor, scalar and vector perturbations. This motivated us to analyze in detail the class of potentials in eq.~\eqref{pot_sin}, which share the preceding property concerning $\mu$ and $\tilde{\mu}$ and afford exact solutions in terms of associated Legendre functions. This allowed us to perform a complete span of the constant mass curves, in order to characterize the stability regions.

Eq.~\eqref{pot_sin} provides a very good approximation for the potential of tensor perturbations, and good ones for those of scalar and vector perturbations in these vacua, as the reader can see in fig.~\ref{fig:exact_vs_sin}. For tensor perturbations $\mu=0$, so that there is a unique potential of the type~\eqref{pot_sin} of the ${\cal A}{\cal A}^\dagger$ form, with $\beta=\frac{1}{2}$, and the agreement between the real potential and eq.~\eqref{pot_sin} is actually striking. The correspondence is still good but less accurate in the other two cases of scalar and vector perturbations, for which $\mu=1$ and $\mu=\frac{3}{8}$, with a relative preference for the $H_-$ choice, to which we shall return shortly.

Using the results of Section~\ref{sec:selfadjoint}, we can now summarize the lessons that can be drawn from the preceding analysis of the potentials~\eqref{pot_sin}.
\begin{itemize}
\item \textbf{Tensor perturbations, $(\mu=0)$.}
    In this case there is a priori an $SL(2,R) \times U(1)$ moduli space associated to the possible boundary conditions, and in particular the boundary of $SL(2,R)$ parametrizes the independent boundary conditions at the two ends, while the $H_+$ and $H_-$ choices coincide. However, as we have seen in the preceding section, all choices of independent boundary conditions lead to instabilities, aside from the special one corresponding to a wavefunction with no logarithmic singularities at the ends. As a result, \emph{a unique boundary condition is compatible with stability, and the resulting spectrum},
    \beq
    M^2 \ = \ \left(\frac{\pi}{z_m}\right)^2 \, n(n+1) \ , \qquad n=0, 1, \ldots \ ,
    \eeq
    \emph{includes a massless graviton as its low--lying mode}, whose normalized ground--state wavefunction~\eqref{zeromodepsi} is
    \beq
\psi_0 \ = \ \sqrt{\frac{\pi}{2\,z_m} \, \sin \left(\frac{\pi\, z}{z_m}\right)} \ .
    \eeq
    To reiterate, the Dudas--Mourad vacuum does lead to a non--vanishing Newton constant in nine dimensions, as shown in~\cite{dm_vacuum}, and it does so while also leaving there a long--range gravitational force.
\item \textbf{Scalar perturbations, $(\mu=1)$.}
In this case the
boundary conditions are fixed, and the issue is optimizing the overall shift. Fig.~\eqref{fig:exact_vs_sin} already indicates that the $H_-$ choice is preferable with respect to $H_+$. However, one can estimate the best choice for the shift $a$, starting from $H_+ + \frac{\pi^2}{z_m^2}\,a$, with the normalized zero--mode wavefunction~\eqref{zeromodepsi},
\beq
\psi_0 \ = \ \sqrt{\frac{3\,\pi}{4\,z_m} }\left[ \sin \left(\frac{\pi\, z}{z_m}\right) \right]^\frac{3}{2} \ ,
\eeq
and demanding that
\beq
\Delta\,m^2 \ = \ \langle \psi_0 | \left[ V_{true}(z) \ - \ V_+(z) \ - \ \frac{\pi^2}{z_m^2}\, a\right] | \psi_0\rangle  \label{pertas}
\eeq
be as small as possible in absolute value. In fact, in this case the first contribution is negligible with respect to the second, and therefore one can choose
\beq
a \ \simeq \ - \ \frac{z_m^2}{\pi^2}\,\langle \psi_0 | V_+(z) | \psi_0\rangle \ = \ \frac{9}{8} \ ,
\eeq
so that the best Legendre potential is
\beq
V \ \simeq \ \left(\frac{\pi}{z_m}\right)^2 \left[ \frac{3}{4 \left[\sin\left(\frac{\pi\,z}{z_m}\right)\right]^2} \ - \ \frac{9}{8} \right] \ = \ V_- \ - \ \frac{7}{8}\, \left(\frac{\pi}{z_m}\right)^2\ .
\eeq
With $H_+$ the spectrum would contain a massless mode, but taking the correction into account our estimate for the scalar spectrum is
   \beq
    M^2 \ \simeq \ \left(\frac{\pi}{z_m}\right)^2 \Big[ n(n+1) \,-\, \frac{7}{8}\Big] \ , \qquad n= 1, \ldots \ ,
    \eeq
and is purely massive. Therefore, the low--lying dilaton mode that emerges in the nine--dimensional effective theory is not a modulus, as in the conventional Kaluza--Klein setting, but it is stabilized. The common goal of stabilizing moduli is thus realized in this simple case.
\item \textbf{Vector perturbations, $\left(\mu=\frac{3}{8}\right)$.}
This sector belongs to the region $0<\mu<1$, and therefore, as for tensor perturbations, there are in principle infinitely many choices of self--adjoint boundary conditions. However as shown in fig.~\ref{fig:instabilities_nu_real}, demanding stability for the $H_+$ Hamiltonian excludes all boundary conditions except for those corresponding to the solution of
\beq
{\cal A}_+^\dagger \, \psi \ = \ 0 \ ,
\eeq
which is
\beq
\psi_0 = \sqrt{\frac{7\,\sqrt{\pi}\,\Gamma\left(\frac{7}{8}\right)}{3\,z_m\,\Gamma\left(\frac{3}{8}\right)}}\, \left[\sin\left(\frac{\pi\,z}{z_m}\right)
\right]^\frac{7}{8}  \label{zeromode_vector}
\eeq
and its excitations.

For $\mu=\frac{3}{8}$ the two options, $H_+$ and $H_-$, are different, and the latter appears preferable by inspection, so that there is a residual moduli space of stable boundary conditions with only massive modes. One can again estimate the best choice for the shift $a$, starting from $H_+ + \frac{\pi^2}{z_m^2}\,a$ and the zero--mode wavefunction~\eqref{zeromode_vector}, so that
\beq
\Delta\,m^2 \ = \ \langle \psi_0 | \Big[ V_{true}(z) \ - \ V_+(z) \ - \ \frac{\pi^2}{z_m^2}\, a\Big] | \psi_0\rangle \ . \label{perta}
\eeq
The best choice for $a$ eliminates this correction, and the result is
\beq
a \ \simeq 0.19 \ + \ \left(\frac{3}{8} \ + \ \frac{1}{2}\right)^2 \ . \label{shift_vector}
\eeq
The alternative option of starting from $H_-$ and the corresponding zero mode wavefunction~\eqref{zeromodepsi} would have resulted in singular integrals.
\begin{figure}[ht]
\centering
\includegraphics[width=50mm]{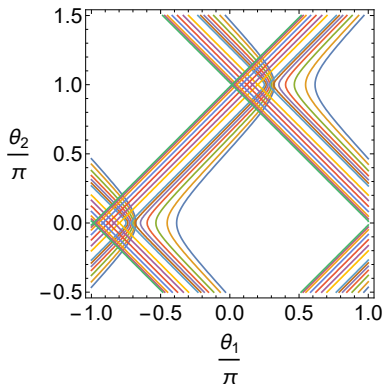}
\caption{\small The wider stability region for vector perturbations implied by the shift in eq.\eqref{shift_vector}.}
\label{fig:wide_vector_stability}
\end{figure}

The corrected stability region for vector perturbations, shown in fig.~\ref{fig:wide_vector_stability}, is wider than the one for $H_-$, which is displayed in the right panel of fig.~\ref{fig:instabilities_nu_real}, since the value obtained for $a$ corresponds to $H_- + 0.2\, \frac{\pi^2}{z_m^2}$. As a result, in fig.~\ref{fig:wide_vector_stability} the only tachyons originate from part of the complex $\nu$ line of eq.~\eqref{nuimaginary}.
\end{itemize}

\section{\sc Conclusions} \label{sec:conclusions}

This work was devoted to identifying the possible self--adjoint boundary conditions for one--dimensional Schr\"odinger systems in finite conformal intervals $0 \leq z \leq z_m$ and their consequences for the nine--dimensional compactifications~\cite{dm_vacuum} of the string models of~\cite{so1616,as95,sugimoto}. We paid special attention to the positivity conditions for the resulting spectra, which translate into the perturbative stability of the vacua.

These compactifications are driven by the tadpole potentials~\eqref{tadpole_intro}, and intervals, rather than the circles of conventional Kaluza--Klein theory, play a prominent role in them and in their generalizations of~\cite{ms21_1,ms21_2}. Logarithmic singularities of the dilaton $\phi$ and the conformal factor $\Omega$ translate into regular singular points of the Schr\"odinger--like differential equations for the modes, which develop double poles at the endpoints of the interval.
These double poles can be characterized by the two parameters $\mu$ and $\tilde{\mu}$ in eq.~\eqref{limiting}, and the resulting wavefunctions have algebraic singularities that depend on them.

As we saw in Section~\ref{sec:sing_pot}, for $\gamma \leq \gamma_c$ only $z_m$ depends on $\gamma$, while the two non--negative parameters $\mu$ and $\tilde{\mu}$ for tensor, vector and scalar modes in eq.~\eqref{limiting} coincide and are independent of it.  These properties reflect a curious correspondence between the asymptotics of these solutions in the presence of tadpole potentials and pairs of Kasner--like solutions in~\cite{ms21_1} that obtain in the absence of them and are mapped into one another by the interchange of $\phi$ with $-\,\phi$.

Characterizing the possible self--adjoint boundary conditions is a crucial prerequisite to addressing the stability of these vacuum solutions with broken supersymmetry of~\cite{dm_vacuum,ms21_1,ms21_2}. Still, while self--adjoint boundary conditions grant the completeness of the corresponding spectra, which is instrumental to describe arbitrary perturbations, they do not provide any clues on the positivity of the resulting real $m^2$ eigenvalues. We have seen that self--adjoint boundary conditions introduce, in general, a number of parameters in the problem, whose values can have crucial effects on vacuum stability. In Section~\ref{sec:selfadjoint} we explored these general boundary conditions for the free theory, which would play a role when considering intervals, along the lines of the Horava--Witten setting of~\cite{hw}, or when the internal space terminates at branes or orientifolds~\cite{pw}. In these cases the boundary conditions resulting in the emergence of instabilities can be linked to the presence of boundary mass terms in the action principle. In the compactifications of interest, the intervals emerge dynamically and the singularities present at the ends make this type of correspondence less direct, but generic self--adjoint boundary conditions do give rise to instabilities. When $\mu$ and $\tilde{\mu}$ are both less than one, the independent choices of boundary conditions are, in fact, in one--to--one correspondence with $SL(2,R) \times U(1)$ matrices, as in the free theory, so that the different options endow these types of compactifications with an additional $AdS_3 \times S^1$ moduli space. Self--adjoint boundary conditions enforced independently at the two ends are a subset of these, correspond to the boundary of $AdS_3$ and are insensitive to the $U(1)$ factor. There are regions of moduli space when one or two tachyonic modes are present, both in the free theory and with the singular potentials of interest. We have explained how, in general, the limiting behaviors of the solutions at the two ends are connected by special $SL(2,R)$ matrices, consistently with the constancy of the Wronskians of the Schr\"odinger operators~\eqref{ham} under scrutiny.  On the other hand, the boundary conditions are unique at ends where $\mu$ or $\tilde{\mu}$ are larger than or equal to one.

For $\gamma \leq \gamma_c$, the equality of $\mu$ and $\tilde{\mu}$ allowed us to rely on a class of exactly solvable potentials related to Legendre functions that can closely approximate the actual potential for tensor perturbations, and reasonably well those for scalar and vector perturbations. They include, in particular, the proper double--pole singularities present in the different cases. We have analyzed the models in detail, and the exact eigenvalue equations allowed us to identify regions of parameter space where tachyonic modes are present and others where the spectrum includes at most massless modes, or is even massive altogether.

For $0< \mu<1$ we have identified pairs of Hamiltonians, $H_\pm$, with $H_
->H_+$, which factor into first--order operators as ${\cal A}_\pm\,{\cal A}_\pm^\dagger$, and we have provided concrete evidence that $H_-$ is stable within wide regions of parameter space while $H_+$ is only stable at special points. Vector perturbations in the ten--dimensional orientifolds correspond to $\mu=\frac{3}{8}$, a value within this range.  Optimizing the correspondence with the actual potential by allowing constant shifts of $H_\pm$, we obtained an improved estimate of the stability region shown in fig.~\ref{fig:wide_vector_stability}, which is wider than the one for $H_-$ displayed in the right panel of fig.~\ref{fig:instabilities_nu_real}.
One can therefore conclude that both purely massive spectra and others also containing massless modes are viable options that can emerge for vectors in the vacua of~\cite{dm_vacuum}, with suitable choices of boundary conditions, consistently with the possible spontaneous breaking of the open--string gauge symmetry.

Our most surprising result concerns gravity. For tensor perturbations $\mu=0$, so that $H_+$ and $H_-$ coincide, and we could show that a long--range gravitational force with a finite value of the nine--dimensional Newton coupling is not only natural for the Dudas--Mourad vacuum of~\cite{dm_vacuum}, as proposed in~\cite{bms}, but it is the only viable option! Other choices of boundary conditions are in principle possible, but they all lead to instabilities here or there, and excluding them we arrived at simple, similar spectra, for tensor and scalar perturbations. The only difference between them is that scalar perturbations lack a massless mode, so that the dilaton is stabilized by the compactification, thus realizing an often sought scenario.

We leave for the future a similar analysis of potentials where $\mu$ and $\tilde{\mu}$ are different, which are relevant for $\gamma>\gamma_c$ and in particular for the $SO(16) \times SO(16)$ model of~\cite{so1616}, but also for the RR modes of the orientifold models of~\cite{as95,sugimoto}.

Summarizing, with the quadratic action principles
\beq
{\cal S} \ = \ \int dz \ {\psi}^\star \left(H \ - \ m^2 \right) \psi
\eeq
underlying the boundary--value problems that we addressed,
if the $\psi$'s are chosen within the self--adjointness domain for $H$, one is led to Schr\"odinger--like equations with the $L^2$ scalar products that we have discussed. In the non--singular case, the boundary conditions~\eqref{bcU} generalize the familiar Neumann and Dirichlet ones. In the singular case, although the wavefunctions are generally non--analytic at the ends of the interval, non--standard boundary conditions can still be characterized in terms of their leading behaviors. The solutions may even satisfy Neumann and Dirichlet conditions at the same time, as we saw for the unique option for $\m=\tilde{\mu}=1$ related to scalar perturbations, while boundary conditions can introduce some moduli of their own, as was the case within the $0 \leq \mu<1$ range, and in particular for the solutions related to vector perturbations. Stability, however, can even reduce a moduli space to a single option, as we saw for tensor perturbations.

What we did not address in detail here is the correspondence between the boundary conditions and possible boundary terms in the action, although we made a cursory comment in this respect, for the free theory, in Section~\ref{sec:large rho}. This correspondence will play a role in the comparison with the current literature on ``dynamical cobordism'', some of which can be found in~\cite{dynamicalcobordism}, and we shall return to it in a future work~\cite{ms23_1}.

It would be interesting to combine the analysis presented in this paper with the methods of ``fake supersymmetry''~\cite{fakesusy}, first considered in this context in~\cite{raucci_23}. The present analysis of boundary conditions will play a prominent role for the Bose modes of the solutions in~\cite{ms21_1} examined in~\cite{ms23_1}, where no strong string--coupling regions are present at the ends of the interval.

\section*{\sc Acknowledgments}
We are grateful to C.~Bachas, G.~Dall'Agata, E.~Dudas, S.~Raucci and A.~Tomasiello for stimulating discussions. AS was supported in part by Scuola Normale, by INFN (IS GSS-Pi) and by the MIUR-PRIN contract 2017CC72MK\_003. JM is grateful to Scuola Normale Superiore for the kind hospitality while this work was in progress. AS is grateful to Universit\'e de Paris Cit\'e and DESY--Hamburg for the kind hospitality, and to the Alexander von Humboldt Foundation for the kind and generous support, while this work was in progress. Finally, we are both grateful to Dr.~M.~Nardelli, who kindly retrieved some mathematical literature.

\end{document}